\documentclass[10pt]{siamltex}
\usepackage{amsmath,amssymb,graphicx,algorithm,algpseudocode,mathdots,color}

\def\fin{\ifmmode{\Large$\diamond$}\else{\unskip\nobreak\hfil
    \penalty50\hskip1em\null\nobreak\hfil{\Large$\diamond$}
    \parfillskip=0pt\finalhyphendemerits=0\endgraf}\fi}

\def\be#1#2\ee{\begin{equation}\label{eq:#1}#2\end{equation}}
\def\req#1{{\rm(\ref{eq:#1})}}
\def\bdm  {\begin{displaymath}}
  \def\edm  {\end{displaymath}}
\def\bdmal{\begin{displaymath}\begin{aligned}}
    \def\edmal{\end{aligned}\end{displaymath}}

\newcommand{\R}{{\mathord{\mathbb R}}}
\newcommand{\N}{{\mathord{\mathbb N}}}
\newcommand{\C}{{\mathord{\mathbb C}}}
\newcommand{\dt}{{\rm d}t}
\newcommand{\ds}{{\rm d}s}

\newcommand{\dW}{{\rm d}W}
\newcommand{\vdot}{\dot{V}}
\newcommand{\Vtilde}{Y}
\newcommand{\Ftilde}{{\widetilde F}}

\newcommand{\gtilde}{{\widetilde\gamma}}
\newcommand{\dVtilde}{{\rm d}Y}

\newcommand{\phihat}{\widehat{\varphi}}

\newcommand{\thsp}{\hspace*{0.1ex}}

\newcommand{\ZZ}{\mathcal{Z}}

\newcommand{\rmi}{\textrm{i}}
\newcommand{\pmax}{\textsl{g}}

\newcommand{\NB}[1]{{\color{blue} #1}}

\newcommand{\MHB}[1]{{\color{red} #1}}
\definecolor{orange}{rgb}{1,0.6,0}
\newcommand{\FS}[1]{{\color{orange} #1}}

\newcommand{\comment}[1]{}

\renewenvironment{matrix}{%
  \matrix@check\matrix\env@matrix
}{%
  \endarray \hskip -\arraycolsep
}

\makeatletter
\newcount\c@MaxMatrixCols \c@MaxMatrixCols=10
\newenvironment{cmatrixc}{%
  \hskip -\arraycolsep\array{*\c@MaxMatrixCols c}%
}{%
  \endarray \hskip -\arraycolsep
}
\makeatother

\newenvironment{cmatrix}{\left[\cmatrixc}{\endmatrix\right]}

\usepackage{mathtools}
\usepackage{algorithm} % for pseudocode
\usepackage{algpseudocode} % for pseudocode
\usepackage{csquotes} % for quotes (\enquote{...})
\usepackage{subfig}
\usepackage{siunitx} % für numerische Spalte in Tabellen
\usepackage{booktabs} % für \toprule, \midrule, \bottomrule
\usepackage{enumitem}

\newcommand{\defin}{\vcentcolon=}
\newcommand{\defininv}{=\vcentcolon}
\DeclarePairedDelimiterXPP{\scalarprodAux}[3]{}{[}{]}{_{#3}}{#1, #2}
\NewDocumentCommand{\scalarprod}{s O{} m m O{}} {\IfBooleanTF{#1}{\scalarprodAux*{#3}{#4}{#5}}{\scalarprodAux[#2]{#3}{#4}{#5}}}
\DeclarePairedDelimiterXPP{\normAux}[2]{}{\lVert}{\rVert}{_{#2}}{#1}
\NewDocumentCommand{\norm}{s O{} m O{}} {\IfBooleanTF{#1}{\normAux*{#3}{#4}} {\normAux[#2]{#3}{#4}}}
\let\T\transpose
\NewDocumentCommand{\identityMatrix}{O{}}{\mathbf{I}_{#1}} % identity matrix

\NewDocumentCommand{\zeroMatrix}{O{}}{\mathbf{0}_{#1}} % zero matrix

\newcommand{\placeholder}{\mkern2mu\cdot\mkern2mu} % placeholder for function argument
\newcommand{\PP}{E} % projection to  V component

\title{Determining extended Markov parameterizations for vector-valued generalized Langevin Equations}

\author{N Bockius\thanks{Institut f\"ur Mathematik, Johannes
    Gutenberg-Universit\"at Mainz, 55099 Mainz, Germany} \and 
    M Braun\thanks{Institut f\"ur Mathematik, Johannes
    Gutenberg-Universit\"at Mainz, 55099 Mainz, Germany} \and 
    K Hofmann\thanks{Institut f\"ur Physik, Johannes
    Gutenberg-Universit\"at Mainz, 55099 Mainz, Germany} \and 
    F Schmid\thanks{Institut f\"ur Physik, Johannes
    Gutenberg-Universit\"at Mainz, 55099 Mainz, Germany} \and 
    M Hanke\thanks{Institut f\"ur Mathematik, Johannes
    Gutenberg-Universit\"at Mainz, 55099 Mainz, Germany
    ({\tt hanke@math.uni-mainz.de})}}

\begin{document}
\maketitle

\begin{abstract}
The generalized Langevin equation is used as a model for various 
coarse-grained physical processes, e.g., the time evolution of the 
velocity of a given larger particle in an implicitly represented solvent,
when the relevant time scales of the dynamics
of the larger particle and the solvent particles are not strictly separated. 
Since this equation involves an integrated history of past
velocities, considerable efforts have been made to approximate this dynamics
by data-driven Markov models, where auxiliary variables are used to compensate
for the memory term. In recent works we have developed two algorithms which
can be used for this purpose, provided the dynamics in question are scalar
processes. Here we extend these algorithms to vector-valued processes. 
As a physical test bed we consider an S-shaped particle sliding on a planar
substrate, which gives rise to a truly two-dimensional velocity process.
The two algorithms provide Markov approximations of this process with 10--20
auxiliary variables and a very accurate fit of the given autocorrelation data
over the entire time interval where these data are non-negligible.
\end{abstract}

\begin{keywords} coarse-graining, model reduction, 
nonsymmetric block Lanczos method, matrix-valued AAA algorithm
\end{keywords}

%\submitto{\JPCM}
%\bigskip 
%\noindent
%Version of \today
%\bigskip
%\ioptwocol

\sloppy

%%%%%%%%%%%%%%%%%%%%%%%%%%%%%%%%%%%%%%%%%%%%%%%%%%%%%%%%%%%
\section{Introduction}
\label{Sec:Intro}
%%%%%%%%%%%%%%%%%%%%%%%%%%%%%%%%%%%%%%%%%%%%%%%%%%%%%%%%%%%
Stochastic differential equations play a central role in many areas of
physics and chemistry~\cite{Risken_book, Pavliotis14}, as they emerge naturally 
in effective descriptions for the motion of particles or collective variables
in a fluctuating surrounding environment. Among these, the generalized
Langevin equation (GLE) provides a particularly flexible
framework~\cite{Zwanzig_book,KTJSV21,Schilling22}.  Unlike simpler stochastic models, such as
the classical Langevin equation that treats fluctuations as instantaneous
and uncorrelated in time, GLEs incorporate memory effects through a
kernel that allows past states of the system to influence its present
behavior, complemented by correlated colored noise.  It arises in
statistical mechanics through projection operator techniques such as the
Mori-Zwanzig formalism~\cite{Zwanzig61,Mori65}, which offers a formally rigorous
unifying framework to connect microscopic dynamics with effective
macroscopic descriptions.  Motivated by this, GLEs are also frequently
proposed as phenomenological models for effective dynamics in slowly
relaxing environments. Applications include, among other, solvent
effects in molecular dynamics, anomalous diffusion and viscoelastic
behavior, chemical reaction dynamics in complex environments, and
descriptions for protein dynamics~\cite{GKC09,Goychuk12,LAD19, Netz24}.

When using GLEs in practice, one faces two main challenges.  The first
is to construct a suitable GLE for a given physical problem. In most
cases, a rigorous derivation from first principles is not possible, and
one instead seeks a GLE that reproduces the relevant statistical
properties of representative trajectories obtained from experiments or
microscopic simulations. Assuming that memory and noise are
coupled to each other -- usually via the so-called fluctuation
dissipation relation, see below~\cite{KTJSV21} -- this reduces to the task of
reconstructing the memory kernel from such data, a problem for which
several strategies have been proposed~\cite{SKTL10, CVR14, CLL14, LBL16, 
SKTL10, MLL16, MLL16a, LLDK17, JHS17, MPS19, GLLB20, WMP20, WPDJ23,TDN24, KWV24}.  

The second challenge
lies in the numerical integration of GLEs.  Direct numerical evaluations
of the memory terms are often computationally demanding, especially when
the memory kernel decays slowly. Moreover, standard efficient
integrators for Markovian equations are no longer applicable for GLEs.
An elegant workaround is the Markovian embedding approach, which maps the
GLE onto a system of coupled Langevin equations without
memory~\cite{FP79,MG83, DBP09, SGTH10, MLL16a, KWV24}, provided that the kernel 
can be decomposed into a suitable sum of exponentials.  
The resulting Markovian system can be
simulated at significantly reduced cost.

In practical applications, however, this two-step workflow -- first
reconstructing a GLE from trajectory data and then mapping it to an
embedded Markovian system -- may be unnecessarily cumbersome and prone to
error. A more direct and efficient strategy is to bypass the
intermediate GLE altogether and construct the embedded Markovian
representation directly from the available
trajectories~\cite{WMP20,BSJSH21}. In earlier work, three of us proposed
such a procedure based on model reduction theory and the positive real
lemma, and demonstrated its effectiveness in the case of one-dimensional
systems~\cite{BSJSH21}.  The present study extends this approach to
multidimensional settings, where several coupled effective degrees of
freedom must be treated simultaneously.

%%%%%%%%%%%%%%%%%%%%%%%%%%%%%%%%%%%%%%%%%%%%%%%%%%%%%%%%%%%
\section{Statement of the problem}
\label{Sec:Problem}
%%%%%%%%%%%%%%%%%%%%%%%%%%%%%%%%%%%%%%%%%%%%%%%%%%%%%%%%%%%
In the following, we will consider GLEs designed to describe the motion
of a single particle in a medium. The GLE has the form
\begin{equation}
\label{eq:GLE}
   M\vdot(t) \,=\, -\!\int_0^t\ds\,\gamma(t-s)V(s) \,+\, F_R(t),
\end{equation}
where $M$ is the mass of the particle and $V=V(t)\in \R^d$ its velocity.
The effect of the medium is encoded in the memory kernel $\gamma(t)$ and
in the fluctuating force $F_R(t)$, which is assumed to be Gaussian
distributed with mean zero and correlations
\begin{equation}
\label{eq:stoch_force}
   C_{F_R}(t-t') \,=\, 
   \langle F_R(t)F_R(t')^T\rangle \,=\, k_B T\,\gamma(t-t')\,
\end{equation}
($k_B T$ is the Boltzmann factor). The so-called fluctuation dissipation
relation (\ref{eq:stoch_force}) can be derived from first principles
using the Mori-Zwanzig projection operator formalism~\cite{Mori65}, and has
been shown to also hold under steady-state non-equilibrium conditions
for free particles with GLEs of type (\ref{eq:GLE}) \cite{JS21,SJS22}. In the
presence of external potentials, its general validity has been
questioned~\cite{GS21, Schilling22}. In that case, Equation (\ref{eq:GLE}) must be
extended to include coupled equations of motion for the position, and
it might be advisable to allow for nonlinear, possibly position-dependent
memory kernels~\cite{VM22}. The treatment of such systems is beyond the
scope of the present work.

%Without loss of generality one may consider the coordinates $V$
%to be chosen in such a way that their covariance is the identity matrix,
%\be{normalization}
%   \langle V(t)V(t)^T \rangle \,=\, I \qquad \mbox{for all $t\in\R$}\,.
%\ee
%For the model problem of Section~\ref{Sec:Phys} this will be the 
%rescaled velocity vector of the center of mass of the anisotropic
%S-shaped% 
%probe particle.

Our intention is to determine numerically a Markovian approximation of the
velocity dynamics by introducing auxiliary variables. More specifically,
we are looking for matrices $A_0\in\R^{N\times N}$, $K\in\R^{d\times d}$,
and $B,C,L\in\R^{N\times d}$, such that the first $d$ components $\Vtilde$
of the stationary solution of the Langevin equation 
\begin{equation}
\label{eq:Langevin}
   {\rm d}\!\begin{cmatrix} \Vtilde \\ Z \end{cmatrix}
   \,=\, \begin{cmatrix}
            \phantom{-}0\phantom{_1} & \!\thsp B^T\\ \!-C & \!A_0 
         \end{cmatrix}
         \begin{cmatrix} \Vtilde \\ Z \end{cmatrix}\dt
       + \begin{cmatrix} \,K\,\\ L \end{cmatrix} \dW\,, \qquad t>0\,,
\end{equation}
with a $d$-dimensional Brownian motion $W$, are close to the rescaled velocity
in that the autocorrelation function $C_\Vtilde$ satisfies
\begin{equation}
\label{eq:interpol}
   C_\Vtilde(\nu\tau) \,\approx\, Y_\nu \,\defin\, \frac{M}{k_BT}\,C_V(\nu\tau)\,, \qquad 
   \nu=0,\dots,2n-1\,.
\end{equation}
In other words, $Y_\nu$ are given snapshots of the rescaled
autocorrelation function $C_V$ of $V$ on an equidistant grid with time step 
$\tau>0$. Take note that, due to this normalization,
\be{normalization}
   Y_0 \,=\, \frac{M}{k_BT}\,\langle V(t)V(t)^T\rangle \,=\, I\,.
\ee

%%%%%%%%%%%%%%%%%%%%%%%%%%%%%%%%%%%%%%%%%%%%%%%%%%%%%%%%%%%
\subsection{Mathematical Approach}
\label{Sec:Math}
%%%%%%%%%%%%%%%%%%%%%%%%%%%%%%%%%%%%%%%%%%%%%%%%%%%%%%%%%%%

In order for \req{Langevin} to have a stationary solution it is necessary
that the block matrix 
\be{A}
   A \,=\, \begin{cmatrix}
              \phantom{-}0\phantom{_1} & \!\thsp B^T\\ \!-C & \!A_0 
           \end{cmatrix}
     \ \in\,\R^{(N+d)\times(N+d)} 
\ee
on the right-hand side of \req{Langevin} is stable, i.e., that all its 
eigenvalues have negative real parts. In this case, cf.~\cite{Pavliotis14},
the autocorrelation function of $\Vtilde$ is given by
\be{f}
   C_\Vtilde(t) \,=\,E^T e^{tA} \Sigma E\qquad \mbox{for $t\geq 0$}\,,
\ee
where $\Sigma$ is the covariance matrix associated with the stationary
solution of \req{Langevin}, and 
\be{E}
   E \,=\, \begin{cmatrix} I \\ 0 \end{cmatrix}
\ee
with a $d\times d$ identity matrix block on top of an $N\times d$ 
zero block. In other words, a multiplication with $E^T$ from the left
projects the solution of \req{Langevin} onto its $\Vtilde$-component. 

It is well-known that $\Sigma$ is the unique solution of the Lyapunov equation
\begin{equation}
\label{eq:Lyapunov}
   A\Sigma \,+\, \Sigma A^T \,=\, 
   - \begin{cmatrix} KK^T & KL^T\, \\ LK^T & LL^T \end{cmatrix},
\end{equation}
and without loss of generality we assume that
\begin{equation}
\label{eq:Sigma}
   \ \Sigma \,=\, \begin{cmatrix} \,I & \,0\, \\ 0 & \Sigma_0 \end{cmatrix}
\end{equation}
with a symmetric positive semidefinite matrix block 
$\Sigma_0\in\R^{N\times N}$ in the lower right corner.
In other words, $\Vtilde(t)$ and $Z(t)$ are taken to be 
uncorrelated with $C_\Vtilde(0) = I = \frac{M}{k_BT}C_V(0)$.
Inserting \req{Sigma} into \req{f} we conclude that
\be{f2}
   C_\Vtilde(t) \,=\, E^Te^{tA} E \,\defininv\, \varphi(t) \qquad 
   \mbox{for $t\geq 0$}\,,
\ee
and we note that 
\[
   C_\Vtilde(t) \,=\, C_\Vtilde(-t)^T \qquad \mbox{for $t<0$}\,.
\]

Eliminating the auxiliary variables $Z$ from the system~\req{Langevin}
it is seen that $\Vtilde$ obeys generalized Langevin dynamics
\be{GLE-tilde}
   \dot{\Vtilde}(t) 
   \,=\, -\!\int_0^t\ds\,\gtilde(t-s)\Vtilde(s) \,+\, \Ftilde_R(t)
\ee
with $\langle\Ftilde_R(t)\Ftilde_R(0)^T\rangle=\gtilde(t)$, where
\be{memory}
   \gtilde(t) \,=\, B^Te^{t\thsp A_0}C\,, \qquad t\geq 0\,,
\ee
cf., e.g., \cite{CBP10}.
When necessary, this function $\gtilde$ can be used for approximating the
memory kernel of the true dynamics~\req{GLE} of the 
underlying system, i.e.,
\bdm
   \gamma(t) \,\approx\, M\,\gtilde(t)\,.
\edm 
% \MHB{Niklas: please check}
Taking the expectation of the product of \req{GLE-tilde}
with $\Vtilde(0)^T$, one further observes 
that the memory kernel $\gtilde$ satisfies the Volterra integral equation
\be{Volterra}
   \dot{C}_\Vtilde(t) \,=\, - \int_0^t \ds\,\gtilde(t-s)C_\Vtilde(s)\,, 
\ee
cf., e.g., \cite{Hanke21,KTJSV21,Kubo66,SKTL10}. 
Note that \req{Volterra} implies that
\be{Vprime}
   \dot{C}_\Vtilde(0+) \,=\, 0\,,
\ee
and that
\be{Vdoubleprime}
   \ddot{C}_\Vtilde(0+) \,=\, -\gtilde(0) 
   \,=\, -\langle \Ftilde_R(0)\Ftilde_R(0)^T\rangle
\ee
is symmetric and negative semidefinite. And finally, differentiating once
again, we conclude that
\be{Vtripleprime}
   \dddot{C}_\Vtilde(0+) - \dddot{C}_\Vtilde(0-)
   \,=\, -\Bigl(\dot{\gtilde}(0+)-\dot{\gtilde}(0-)\Bigr)\,,
\ee
and the right-hand side is positive semidefinite, because $\gtilde$ is
the autocorrelation function of the random force term in \req{GLE-tilde},
and hence, is a function of positive type, cf.~\cite{Pavliotis14}.

Assuming $A$ to be diagonalizable the projected matrix exponential $\varphi$ of 
\req{f2} can be written as a finite Prony series
\begin{equation}
\label{eq:Prony}
   \varphi(t) \,=\, \sum_{j=1}^{p} \Gamma_j e^{\lambda_j t}\,, \qquad
   t\geq 0\,,
\end{equation}
where the exponents $\lambda_j$ are the eigenvalues of $A$ 
and $\Gamma_j\in\R^{d\times d}$ are appropriate coefficient matrices. 
Accordingly, in order to satisfy \req{interpol} one has to select a 
reasonable number $m\in\N$ of appropriate exponents $\lambda_j$ with 
negative real parts and associated $d\times d$ coefficient 
matrices $\Gamma_j$, such that
\be{approx}
   \varphi(\nu\tau) \,=\, \sum_{j=1}^p \Gamma_j e^{\nu\tau\lambda_j} 
   \,\approx\, Y_\nu\,, \qquad \nu=0,1,\dots,2n-1\,.
\ee
However, not every Prony series $\varphi$ of this form is admissible as an
autocorrelation function, since Bochner's theorem (cf.~\cite{Pavliotis14})
states that the Fourier transform of every autocorrelation function must be
nonnegative semidefinite for every frequency $\omega\in\R$. 

Unfortunately, it is an unresolved problem to find simple suitable conditions 
on the coefficients $\lambda_j$ and $\Gamma_j$ of \req{approx}, such that the
Fourier transform $\phihat$ is nonnegative semidefinite. The good
news, on the other hand, is that it often suffices to make sure
that $\varphi$ satisfies the two algebraic conditions which must be 
fulfilled by 
$C_\Vtilde$ according to \req{Vprime} and \req{Vdoubleprime}, namely that
\be{fprime}
   \dot{\varphi}(0) \,=\, \sum_{j=1}^p \lambda_j \Gamma_j \,=\, 0 
   \quad \mbox{and} \quad
   \ddot{\varphi}(0) \,=\, \sum_{j=1}^p \lambda_j^2 \Gamma_j \ \ 
   \mbox{is symmetric}\,,
\ee
to achieve that $\phihat$ is positive semidefinite. Take note that we also
want to have
\be{f0}
   \varphi(0) \,=\, \sum_{j=1}^p \Gamma_j \,=\, C_\Vtilde(0) \,=\, I\,,
\ee
because $C_V(0)=I$ due to our normalization \req{normalization}.

%%%%%%%%%%%%%%%%%%%%%%%%%%%%%%%%%%%%%%%%%%%%%%%%%%%%%%%%%%%
\subsection{A physical model system}
\label{Sec:Phys}
%%%%%%%%%%%%%%%%%%%%%%%%%%%%%%%%%%%%%%%%%%%%%%%%%%%%%%%%%%%

As a test case for our numerical algorithms, we consider the
two-dimensional motion of a rigid S-shaped particle sliding on a
planar substrate and immersed in a bath of $N_{\mathrm{L}}$ spherical
particles of mass $m$ and diameter $\sigma$. The S-shaped particle
undergoes translational motion, but does not rotate. This system is
deliberately chosen for its low symmetry: Not only is the diffusion
tensor anisotropic, but the different velocity components are also
coupled in a nontrivial manner that cannot be eliminated by a simple
axis transformation.  Physically, the rotation of probe particles
could be prevented, e.g., by applying external orienting fields. 

%, immersed in an implicit solvent. 
%The latter is modeled by a Markovian Langevin thermostat, which couples the 
%Langevin particles to a thermal bath of fixed temperature $T$. 
%vorher: temperature $k_BT$
The state of the bath particles is described by their positions $r_i$ and 
velocities $v_i$, and they interact with each other via a
Weeks-Chandler-Anderson (WCA) potential,
\begin{align}
\label{eq:WCA}
    U_\text{WCA}(r_{ij})\,=\,\begin{cases}
        4 \epsilon \left( \left( \dfrac{\sigma}{r_{ij}} \right)^{12} - \left( \dfrac{\sigma}{r_{ij}} \right)^{6} \right) + \epsilon\,, & r_{ij} \leq \sqrt[6]{2}\, \sigma\,, \\[3ex]
        0\,, & r_{ij} > \sqrt[6]{2}\,\sigma \,,
    \end{cases}
\end{align}
where 
%$\epsilon$ is set to unity for simplicity, and 
$r_{ij}$ is the distance between the two interacting particles.
Furthermore, they are coupled to a Langevin thermostat, which mimics
the friction with the substrate. Introducing the implicit substrate
has the advantage of cutting off long-range hydrodynamic interactions, 
which decay only logarithmically in two dimensions and give rise to
pronounced finite size effects and hydrodynamic backflows. Dealing
with such complications is beyond the scope of the present work.
The equations of motion for the $i^\mathrm{th}$ Langevin particle are thus
\begin{align}
    \label{eq:vv}
    m \dot{v}_i &\,=\, -\gamma v_i + \xi_i - \sum_{j\neq i} \nabla U_\text{WCA}(r_{ij})\,, \\
    \label{eq:rr}
    \dot{r}_i &\,=\, v_i\,,
\end{align}
with $\gamma$ denoting a translational friction coefficient.
%according to Stokes and the viscosity $\eta$ of the solvent. 
The random force vector
$\xi$ is defined as Gaussian white noise, with vanishing mean and
autocorrelation
\begin{align}
    C_\xi(t) \,=\, 2 k_\text{B}T \gamma\, \delta(t) I\,,
    \qquad t\in\R\,.
\end{align}

The S-shaped probe particle is taken to be a rigid chain of
$N_\mathrm{s}$ beads with mass $m_\mathrm{s}$ and diameter
$d_\mathrm{s}=\sigma$. Thus, the total mass of the S-shaped probe
results in $M=N_\mathrm{s}m_\mathrm{s}$. The state of the
S-shaped particle is described by the position $R$ and
velocity $V$ of the center of mass. The bead particles
interact with the bath particles with the same WCA potential, Equation
(\ref{eq:WCA}), but they are not coupled to a thermostat, i.e., they
can slide freely on the substrate. Thus, we focus on the effective
dynamics arising from the interaction with the bath particles. 
The equations governing the motion of the S-shaped probe are
\begin{align}
    \label{eq:vv_s}
    M \dot{V} &\,
      =\, - \sum_{i,k} \nabla_k U_\text{WCA}(r_{ik})\,, \\
    \label{eq:rr_s}
    \dot{R} &\,=\, V\,,
\end{align}
where the sum runs over all pairs of bath particles $i$ and bead
particles $k$. Since rotational motion is prevented, 
the S-shaped probe maintains a fixed orientation. 
Due to the planar substrate, the
motion of the Langevin particles and the S-shaped probe is confined to
a two-dimensional plane. All quantities will be given in units of the
length $\sigma$, the thermal energy $k_\mathrm{B}T$, and a \enquote{natural}
time unit $t_0=\sqrt{m\sigma^2/k_\mathrm{B}T}$, which corresponds to
the inertial time scale of bath particles.  
We set $\gamma = 3\pi k_\mathrm{B}Tt_0$, which represents the friction coefficient
of a spherical particle in a Newtonian fluid with viscosity $\eta =
1\,k_\mathrm{B}Tt_0/\sigma$ (i.e. $\gamma=3\pi\eta\sigma$). In the
WCA potential (Eq.~\eqref{eq:WCA}), we choose $\epsilon =
k_\mathrm{B}T$.

%%%%%%%%%%%%%%%%%%%%%%%%%%%%%%%%%%%%%%%%%%%%%%%%%%%%%%%%%%%
\section{Methods}
\label{Sec:Methods}

%%%%%%%%%%%%%%%%%%%%%%%%%%%%%%%%%%%%%%%%%%%%%%%%%%%%%%%%%%%
\subsection{Simulation details}
\label{subsec:data}

\begin{figure}
    \centering
    \includegraphics[width=0.5\linewidth]{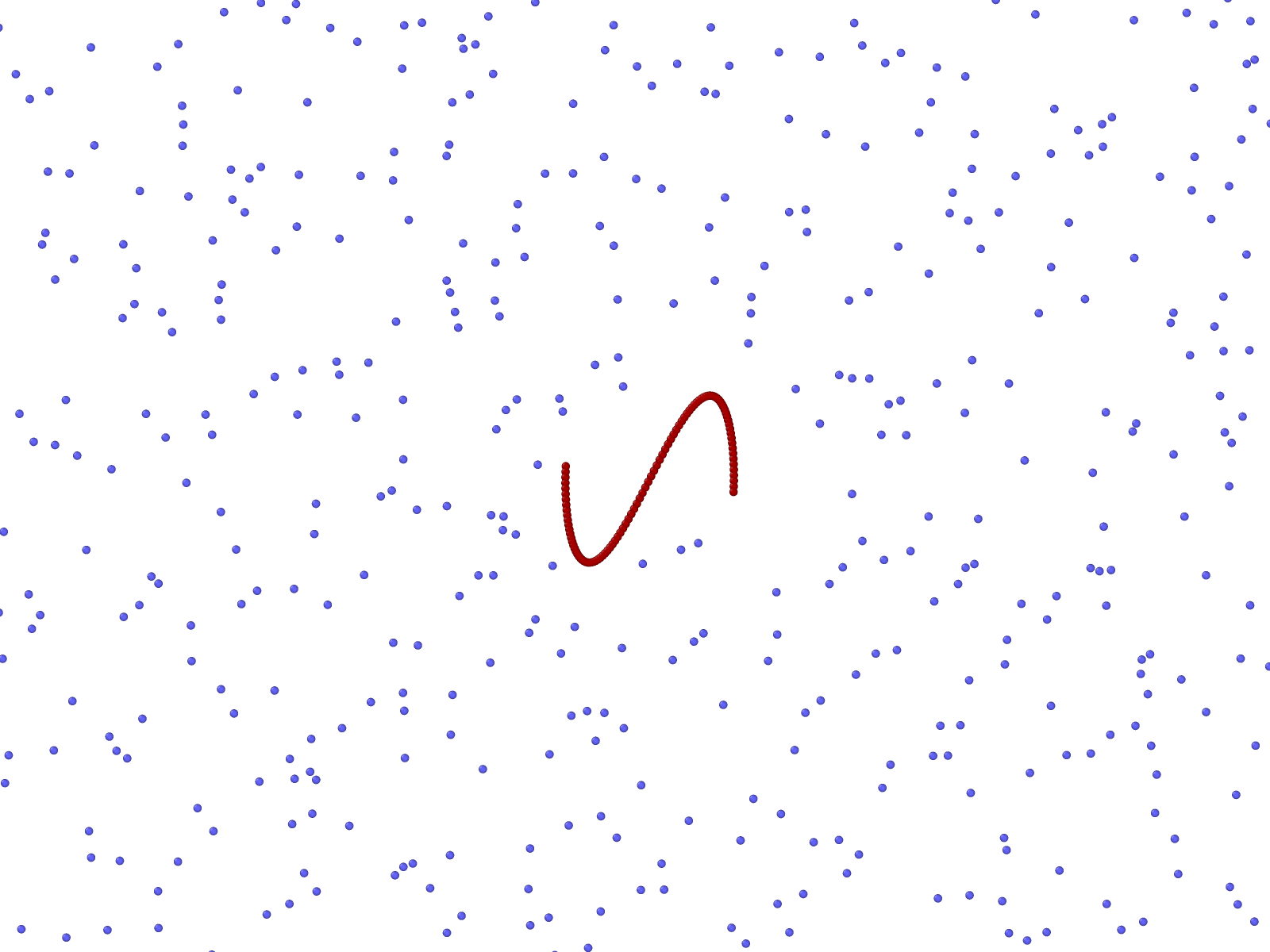}
    \caption{Snapshot of the simulated ensemble.}
    \label{fig:Sshape}
\end{figure}

The trajectories of the molecular dynamics simulations are generated
according to the Langevin equation of
motion~\eqref{eq:vv}-\eqref{eq:rr_s} modelling a canonical ensemble,
as implemented in the HOOMD-blue package \cite{HOOMD}. Each bath
particle has a mass $m=1\,k_\mathrm{B}Tt_o^2/\sigma^2$. The S-shaped
probe consists of $N_\mathrm{s}=100$ beads each with a mass of
$m_\mathrm{s}=0.03\,m$. The positions $r_k$ of the constituent beads,
which make up the particular shape displayed in
Figure~\ref{fig:Sshape}, are calculated by $r_k=R +10\cdot(\sin{t_k},
\sin{2t_k})$, where $R$ is the position of the center of mass of the probe and $t_k$
is evenly spaced in the interval $[-\pi+1.5,\pi-1.5]$. 
The bath consists of $N_L =5000$ particles in a total simulation
area of size $43.113 \times 43.113 \,\sigma^2$ with periodic boundary
conditions.
%The thermal bath temperature is set to $1\,k_\text{B}T$. 
Each simulation run is
split in an equilibration period of $10^5$ time steps and a production
run of $10^8$ time steps with a step size of $\Delta t=0.001\,t_0$.

%%%%%%%%%%%%%%%%%%%%%%%%%%%%%%%%%
\subsection{A block version of Prony's method}
\label{Subsec:Prony}
%%%%%%%%%%%%%%%%%%%%%%%%%%%%%%%%%
The first algorithm that we consider for constructing a suitable 
Langevin system~\req{Langevin} generalizes the method from
\cite{BSJSH21} to vector-valued processes.
It attempts to construct a finite Prony series~\req{Prony}
which interpolates the given time samples,
\be{expinterpol}
   \varphi(\nu\tau) \,=\, Y_\nu\,, \qquad \nu=0,\dots,2n-1\,,
\ee
i.e., \req{approx} holds true with equality signs.

We use these time samples to define the block Hankel matrix
\begin{equation} \label{eq:hankel-matrix}
	\mathcal{H} \,=\, \begin{cmatrix}
		Y_0 & Y_1 & \cdots & Y_n \\
		Y_1 &  & \iddots & \vdots \\
%		Y_2 & Y_3 & Y_4 & \ldots \\
	    \vdots & \iddots & & Y_{2n-1} \\
        Y_n & \dots & Y_{2n-1}\!\!\!\! & 0 
	\end{cmatrix}
\end{equation}
and the induced bilinear form
\begin{equation} \label{eq:lanczos-bilinear-form-matrix-product-shape}
	\scalarprod{\psi}{\phi}[\mathcal{H}] \,=\, \T{a}\mathcal{H}b
\end{equation}
for polynomials
\begin{equation*}
	\psi(x) = \sum\limits_{i=0}^{(n+1)d-1} a_{i}x^i \quad \text{and} \quad 
    \phi(x) = \sum\limits_{i=0}^{(n+1)d-1} b_{i}x^i
\end{equation*}
of degree $(n+1)d-1$ with coefficient vectors $a = \T{[a_0, a_1, \ldots]}$ 
and $b = \T{[b_0, b_1, \ldots]}$ in $\R^{(n+1)d}$.
In contrast to the scalar case the bilinear form 
\req{lanczos-bilinear-form-matrix-product-shape} does not correspond to a
moment functional, and since the matrix $\mathcal{H}$ does not need to be
symmetric, the bilinear form also fails to be symmetric in general. Instead
of the symmetric Lanczos process, which is behind the scalar Prony method 
utilized in \cite{BSJSH21}, we therefore have to resort to the nonsymmetric 
(block) Lanczos process, which determines two bases $\{\psi_k\}$
and $\{\phi_k\}$ of polynomials which are biorthogonal with respect to 
$\scalarprod{\placeholder}{\placeholder}[\mathcal{H}]$, i.e.
\begin{equation*}
	\Bigl|\scalarprod{\psi_k}{\phi_l}[\mathcal{H}]\Bigr| \,=\, \delta_{kl}
    \qquad \text{for} \ k,l=0,\dots,2n-1\,.
\end{equation*}

%Note that $\mathcal{H}$ is only partially defined (up to the entries given by 
%$Y_{2n-1})$ and consequently 
%$\scalarprod{\psi}{\phi}[\mathcal{H}]$ is only defined for polynomials whose 
%degrees are not too large, 
%i.e.\ all entries in $\mathcal{H}$ which are not defined by to $Y_0, \ldots, %Y_{2n-1}$ have to be multiplied 
%by zero.

%By virtue of \req{normalization} we can always assume that $Y_0=I$.
%Let us first assume $Y_0 = \idmat[d]$. 
%(Otherwise we multiply all $Y_k$ by $Y_0^{-1}$, see below.) 
To be specific, we apply the algorithm from \cite{Freund01} to construct these two bases, see Appendix~\ref{Subsec:Lanczos}. 
%two finite sequences $\phi = (\phi_k)_{k=0, \ldots, n-1}$, $\psi = (\psi_k)_{k=0, \ldots, n-1}$ 
%(In contrast to the one-dimensional case from \cite{BSJSH21}, $\scalarprod{\placeholder}{\placeholder}
%[\mathcal{H}]$ is in general not symmetric as $\mathcal{H}$ is not symmetric, hence we need two sequences %instead of a single one.) 
This method proceeds similarly to the Gram-Schmidt algorithm by successively orthogonalizing certain 
basis polynomials of increasing degrees against the previous ones. 
But we note that the algorithm may fail because the bilinear form is
no inner product; such a potential failure has also been a threat in the 
scalar case, 
but since our polynomial degrees are of moderate size, we have never 
encountered this problem in our computations.

The corresponding recursion coefficients make up the entries of a 
banded matrix $T\in\R^{dn\times dn}$ with upper and lower bandwidth $d$, 
which satisfies
\begin{equation} \label{eq:J-interpolation-property}
	\T{\PP}T^k\PP = Y_k \qquad \text{for $k=0, \ldots, 2n-1$}\,,
\end{equation}
with $E$ of \req{E}. Consequently, the function $\varphi$ defined in \req{f2},
using
\be{logJ}
   A \,=\, \frac{1}{\tau}\,\log T\,,
\ee
where $\log T$ denotes the matrix logarithm of $T$
(cf.~Higham~\cite{Higham08}), solves the interpolation problem 
\eqref{eq:expinterpol}, and hence, the matrix in \req{logJ} 
is our candidate for the system matrix $A$ in \eqref{eq:A}. 

As in \cite{BSJSH21} the resulting Prony series may contain exponents 
$\lambda$ with nonnegative real parts.
We call the corresponding terms \enquote{spurious} exponentials since they do not converge to $0$ and are thus unphysical. 
Since our data come from a physical model,
the corresponding coefficients of the Prony series are small compared to the other ones unless the noise is too large. Eliminating these terms 
will therefore have a negligible impact on the values of the Prony series 
for $t \leq (2n-1)\tau$. We therefore remove the corresponding eigenvalues 
from $T$ and $A$; the technical details are postponed to 
Appendix~\ref{Subsec:Lanczos-eigenvalues}.

Another issue, which we have to take care of, are negative real eigenvalues 
of $T$, for otherwise $A$ of \req{logJ} will not be a real-valued matrix,
and the Prony series $\varphi$ will not be a real-valued function. 
Again we already encountered this problem in the scalar case, 
and we fix it here in the very same way as in \cite[Appendix~B.2]{BSJSH21}.
%compare Appendix~\ref{sec:app-eigenvalues}.
%Simply setting $\ref{sec:app-algorithm}. A \defin \frac{1}{\tau}\log J$ would 
%lead to a complex eigenvalue of $A$ whose complex conjugate is no eigenvalue, 
%leading to a non-real VACF approximation. 
%We can solve this problem by \enquote{duplicating} $\mu$ such that it is 
%represented by two complex conjugate eigenvalues in $A$. 
%This does not modify the resulting Prony series. We again postpone the details
%to \ref{sec:app-eigenvalues} for technical details.

So far we haven't discussed the two side conditions in \req{fprime}.
In the scalar case, where $d=1$, the second condition in \req{fprime}
is void, so that \req{fprime} amounts to the single scalar constraint
\be{fprime-scalar}
   \dot{\varphi}(0) \,=\, 0\,.
\ee
In \cite{BSJSH21} this property has been achieved by taking 
the (scalar) function value $Y_1$ for the first nonzero grid point  
as a free parameter, and fit this parameter in an outer Newton iteration 
to achieve the constraint \req{fprime-scalar} for the solution of the 
interpolation problem~\req{expinterpol}. In the vector-valued case 
the situation is significantly more difficult, because \req{fprime} now 
consists of $3d^2/2-d/2$ scalar equations; already when $d=2$ these are five 
scalar equations to be fulfilled. The entries of $Y_1$ no longer provide
enough parameters to fit, and on top of that we have experienced numerically 
that the higher dimensional Newton scheme is prone to deliver inappropriate
Prony series. 

For this reason it is with a heavy heart that we wave the 
side constraints~\req{fprime} when using the above approach. 
Note that it follows from \req{f2} that
\[
   \dot{\varphi}(0) \,=\, E^TAE
\]
is the upper left block entry of the system matrix $A$. Accordingly, 
if $D\defin-\dot\varphi(0)\neq 0$ for the chosen Prony series,
then the upper left block of the final system matrix will be nonzero.
Instead of \req{Langevin} the Langevin dynamics then assumes the form
\begin{equation}
\label{eq:Langevin-nonzero}
   {\rm d}\!\begin{cmatrix} \Vtilde \\ Z \end{cmatrix}
   \,=\, \begin{cmatrix}
            \!-D & \!\thsp B^T\\ \!-C & \!A_0 
         \end{cmatrix}
         \begin{cmatrix} \Vtilde \\ Z \end{cmatrix}\dt
       + \begin{cmatrix} \,K\,\\ L \end{cmatrix} \dW\,,
\end{equation}
and in order to have a stationary solution with some covariance matrix $\Sigma$
as in \req{Sigma} it must be required that $D+D^T$ is positive semidefinite 
with
\be{D-K}
   D+D^T \,=\, KK^T\,;
\ee
this is a consequence of the Lyapunov equation~\req{Lyapunov}.
As shown in \cite{CBP10} the corresponding component $Y$ then satisfies the 
generalized Langevin dynamics
\be{GLE-tilde-friction}
   \dVtilde(t) 
   \,=\, \Bigl(-D\Vtilde(t) \,-\!\int_0^t\ds\,\gtilde(t-s)\Vtilde(s)
                \,+\, \Ftilde_R(t)\Bigr)\dt
         \,+\, K\dW(t) 
\ee
with an extra damping and associated random force term. 
Further, the autocorrelation function of $\Ftilde_R$ now takes the form
\[
   \langle \Ftilde_R(t)\Ftilde_R(0)^T \rangle
   \,=\, \gtilde(t) \,-\, B^Te^{|t|A_0}LK^T\,.
\]
Fortunately the model~\req{GLE-tilde-friction} 
is also suitable from a physical point of view, with $D$ being associated with
an instantaneous damping coefficient.

% To make sure that $D+D^T$ is actually positive semidefinite, 
% we modified the aforementioned technique from \cite{BSJSH21} and 
% shift the second data point $Y_1$ downwards, when necessary;
% to be specific we replace $Y_1$ by $Y_1-10^{-3}I$ and redo the overall
% computation, until $D+D^T$ eventually becomes positive semidefinite.

%However, for the data considered in Section~\ref{Sec:Results}, the matrix
%$D$ failed to be positive semidefinite 
%method often produced a matrix $A$ with an upper left block $-D$ such that 
%$D+D^T$ was not positive semi-definite, which implies that the resulting 
%approximating  $\varphi$ from \eqref{eq:expinterpol} is no valid 
%autocorrelation function since \eqref{eq:D-K} has no solution. 
%Here we solved this problem by successively subtracting a small multiple of 
%the identity matrix (more precisely $0.001\identityMatrix$) from the second 
%data point $Y_1$ and repeating the method with these modified input data until
%it succeeded. 
%When using other data where t  he corresponding GLE is of the type 
%\eqref{eq:GLE-tilde-friction} with $D+D^T$ positive definite, this modification 
%is often unnecessary.

Below we will denote the above algorithm for setting up the 
Langevin system~\req{Langevin-nonzero},
respectively the Prony series $\varphi$, by Method~A for short.

%%%%%%%%%%%%%%%%%%%%%%%%%%%%%%%%%
\subsection{A method based on rational approximation}
\label{Subsec:rat_app}
%%%%%%%%%%%%%%%%%%%%%%%%%%%%%%%%%
The introduction of the additional instantaneous damping coefficient 
in \req{GLE-tilde-friction} can be avoided with a different approach, 
which has originally been suggested in \cite{Hanke24} for scalar-valued
stochastic processes. Here we provide an extension of this method to 
vector-valued generalized Langevin dynamics, later referred to as Method~B.

Following \cite{Hanke24} we introduce the generating function
\be{gen-Func}
    F(z) = \sum_{\nu=0}^{\infty} Y_\nu z^{-\nu-1}
\ee
of all equidistant snapshots $Y_\nu$. 
This function is analytic in the exterior of the unit disk.
%and extends 
%continuously onto the unit circle, when the snapshots are absolutely summable.
Under the assumption that \req{approx} holds true with equality for all 
$\nu\in\N_0$, the generating function can be rewritten as
\be{rat-app}
    F(z) \,=\, \sum_{\nu=0}^\infty 
               \sum_{j=1}^p \Gamma_j e^{\nu\tau\lambda_j}z^{-\nu-1} 
         \,=\, \sum_{j=1}^p \frac{\Gamma_j}{z} 
               \sum_{\nu=0}^\infty (e^{\tau\lambda_j}/z)^\nu 
         \,=\, \sum_{j=1}^p \frac{\Gamma_j}{z - z_j}
\ee
with
\[
    z_j = e^{\tau\lambda_j}\,, \quad j=1,\ldots,p \,.
\]
We therefore proceed by approximating $F$ by a rational function with real
coefficients in the exterior of the unit disk in order to use the poles of this approximation for the construction of a suitable Prony series $\varphi$.
Since $F$ is a matrix-valued function, we use a matrix-valued variant of the AAA algorithm for computing this rational approximation, 
cf.~Appendix~\ref{Subsec:AAA} for details.

For every pole $z$ of this rational approximation with $|z| < 1$, 
we include the exponent(s)
\[
    \lambda = \begin{cases}
        \displaystyle \frac{1}{\tau} \log z\,, & z\notin\R^{-}\,,\\[2ex]
        \displaystyle \frac{1}{\tau} \log |z| \pm \frac{\pi}{\tau}\,\rmi\,, & z\in\R^{-}\,,
    \end{cases}
\]
in the Prony series, whereas poles with $|z|\ge 1$ are discarded;
the latter may occur as artefacts of the rational approximation.
Since our rational approximation has real coefficients, the resulting
exponents $\lambda$ are either real or appear in complex conjugate pairs.
It then remains to choose appropriate coefficient matrices 
$\Gamma_j\in\C^{d\times d}$ for the Prony series~\req{Prony}.
In view of our goal~\req{approx} and the aforementioned side 
constraints~\req{fprime} and \req{f0} we consider the equality constrained least-squares problem
\be{LSQ}
\begin{array}{l}
    {\displaystyle 
    \textrm{minimize} \quad \sum_{\nu=0}^{2n-1}\left\Vert \sum_{j=1}^p \Gamma_j e^{\lambda_j \tau\nu}  - Y_\nu \right\Vert_F^2}\\[4ex]
    {\displaystyle\textrm{subject to}\quad \sum_{j=1}^p \Gamma_j=I\,, \quad
    \sum_{j=1}^p \lambda_j\Gamma_j=0\,, \quad \mbox{and} \quad \sum_{j=1}^p \lambda_j^2 (\Gamma_j-\Gamma_j^T) = 0}\,,
\end{array}
\ee
for this purpose, where $\|\,\cdot\,\|_F$ refers to the Frobenius norm. 
This is a well-understood problem in numerical linear algebra
which can be transformed into the solution of a linear system of equations,
cf.\ Bj\"orck~\cite{Bjor24}. 
Moreover, it can be shown that the matrix coefficients $\Gamma_j$ corresponding
to real exponents $\lambda_j$ have real entries, and that the matrices
$\Gamma_j$ corresponding to complex conjugate pairs of exponents are also
complex conjugates of each other. Accordingly, the resulting
Prony series $\varphi$ is a real-valued function.

To achieve the desired representation
\be{desire}
   \varphi(t) \,=\, E^T e^{tA}E
\ee
of this function, we have to assemble a matrix $A$ with 
eigenvalues $\lambda_j$ and suitable eigenvectors. This somewhat technical
detail is worked out in Appendix~\ref{Subsec:sysmat}.

%%%%%%%%%%%%%%%%%%%%%%%%%%%%%%%%%
\subsection{The Positive Real Lemma}
\label{Subsec:PRL}
%%%%%%%%%%%%%%%%%%%%%%%%%%%%%%%%%
Once the system matrix $A$ of \req{A} has been determined by either of the
two approaches from Sections~\ref{Subsec:Prony} or \ref{Subsec:rat_app},
it remains to construct the matrix blocks $K$ and $L$, which steer the 
Brownian motion in \req{Langevin}, and to determine the associated covariance 
matrix $\Sigma$.
Note that when inserting \req{A} and \req{Sigma} into the 
Lyapunov equation~\req{Lyapunov}, we see that the latter is equivalent to the system
\begin{equation}
\label{eq:singLure}
   K \,=\, 0\,, \quad A_0\Sigma_0+\Sigma_0A_0^T \,=\, -LL^T\,,\quad
   \Sigma_0 B \,=\, C\,.
\end{equation}
These are the so-called (singular) Lur'e equations.
Likewise, if an instantaneous damping term is added to the system 
as in \req{GLE-tilde-friction}
then the corresponding equations become the (regular) Lur'e system
\be{Lure}
   D + D^T \,=\, KK^T\,, \quad
   A_0\Sigma_0+\Sigma_0A_0^T \,=\, -LL^T\,,\quad
   C - \Sigma_0 B \,=\, LK^T\,.   
\ee

It is known, cf.~\cite{AV73}, that for our input matrices $A_0, B, C$ -- and $D$
in \req{Lure} --
the equations~\req{singLure} and \req{Lure},
respectively, have a solution consisting of the matrices
$K\in\R^{d\times d}$, $L\in\R^{N\times d}$, and $\Sigma_0\in\R^{N\times N}$
(the latter being symmetric and positive definite) if and only if
the Prony series $\varphi$ (with its proper extension to the negative time axis)
satisfies the requirements of Bochner's theorem. In other words, 
%if $\varphi$
%qualifies as an autocorrelation function, then the respective Lur'e system
%can be solved for the corresponding factors of the driving force term for %the Langevin equation~\req{Langevin} and the associated covariance matrix 
%of the auxiliary variables of its stationary solution.
%And vice versa: 
if the Lur'e equations fail to have a solution
$(K,L,\Sigma_0)$, then there is no stochastic process whose autocorrelation
function coincides with the Prony series~\req{Prony}. 

The numerical solution of the Lur'e equations is a bit tricky, 
cf.~\cite{WSW90}. We therefore provide in Appendix~\ref{App:ZZZ}
a sketch of the corresponding algorithm under the generic assumptions that
\begin{subequations}
\label{eq:BCBAC}
\be{BC}
   \ddot\varphi(0+) \,=\, B^TC \quad \mbox{is positive definite}
\ee
and
\be{BAC}
   \dddot\varphi(0+) + \dddot\varphi(0+)^T \,=\, - B^TA_0C - C^TA_0^TB
   \quad \mbox{is positive definite}
\ee
\end{subequations}
in the case of the singular Lur'e equations~\req{singLure}, 
compare~\req{Vdoubleprime} and \req{Vtripleprime}, and that
\be{Dnonzero}
   \dot\varphi(0+) + \dot\varphi(0+)^T \,=\, -(D+D^T) \quad 
   \mbox{is negative definite}
\ee
in the regular case~\req{Lure}, respectively.

%%%%%%%%%%%%%%%%%%%%%%%%%%%%%%%%%%%%%%%%%%%%%%%%%%%%%%%%%%%
\section{Numerical results}
\label{Sec:Results}
%%%%%%%%%%%%%%%%%%%%%%%%%%%%%%%%%%%%%%%%%%%%%%%%%%%%%%%%%%%

\begin{figure}
    \centering
    \includegraphics[width=\linewidth]{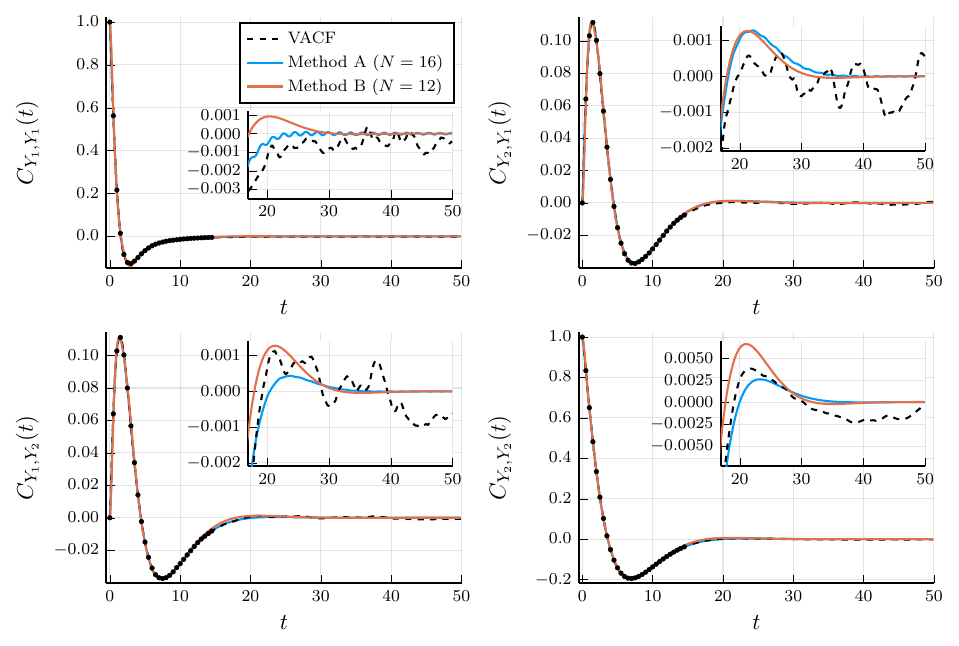}
    \caption{Rescaled velocity autocorrelation function (VACF) 
%    \textcolor{red}{in units of $\sigma^2/t_0^2$}
    versus time $t$ in units of $t_0$ and the
    two approximations: Method~A is the block Prony method from Section~\ref{Subsec:Prony} and Method~B is the one from Section~\ref{Subsec:rat_app} based on rational approximation. 
    $N$ is the number of auxiliary variables of the respective
    Langevin system. 
    Every panel shows the scalar graphs corresponding to the
    respective entry of the (matrix-valued) function.}
    \label{fig:example1}
\end{figure}

We now present numerical examples for 
%the autocorrelation function of 
the two-dimensional velocity (i.e., $d=2$) of the S-shaped particle
described in Section \ref{Sec:Phys}.
%Since the covariance matrix of the corresponding stationary 
%velocities agrees well with a positive multiple of the identity matrix,
%we can simply rescale the given autocorrelation data to achieve the
%normalization~\req{normalization}. 
%Otherwise, one could compute the Cholesky factorization of this covariance 
%matrix in a preliminary step, and 
%multiply the original velocity variables by the inverse Cholesky factor to 
%obtain~\req{normalization}. 
As mentioned before our target is the autocorrelation function 
of the rescaled variables, cf.~\req{interpol}, which is shown as black dashed 
line in Figure~\ref{fig:example1}.

%%%%%%%%%%%%%%%%%%%%%%%%%%%%%%%%%%%%%%%%%%%%%%%%%%%%%%%%%%%%
%\subsection{Example 5.1}
%\label{Subsec:Ex4.2}
%%%%%%%%%%%%%%%%%%%%%%%%%%%%%%%%%%%%%%%%%%%%%%%%%%%%%%%%%%%%
In our first test case we use
%The data we have consist of 600 time samples in total from which we used 
$2n = 30$ samples of these data with a grid spacing of 
$\tau = 0.5\, t_0$, highlighted as black dots in this graph;
recall that $t_0$ is the \enquote{natural} time unit introduced in 
Section~\ref{Sec:Phys}.

The system matrix obtained by Method~A (of Section~\ref{Subsec:Prony})
is of size $30\times 30$, because $dn=30$, too, in the two-dimensional case.
Since we ignore the side constraints~\req{fprime}, the $(1,1)$-block $-D$ 
of $A$ is nonzero; compare~\req{Langevin-nonzero}. 
%However, in order that $D+D^T$ is positive semidefinite, we had to replace the second data sample $Y_2$ by $Y_2-3\cdot 10^{-3}I$. 
This system matrix has $13$ spurious eigenvalues and one 
negative real eigenvalue. 
The elimination of the spurious eigenvalues and duplication of the negative real eigenvalue results in a $18\times18$ system matrix $A$, which corresponds to $N=18-d=16$ auxiliary variables.
%and the spectral norm of the resulting 
%(artificial) damping coefficient in \req{GLE-tilde-friction} is given 
%by $0.84$ in the original physical units, i.e., $1/t_0$.
The corresponding regular Lur'e equations are solvable, which shows that the 
computed Prony series is a valid autocorrelation function. 

\comment{\FS{FS: I would recommend not to call this "drift", because this
creates the wrong association in a physicist reader.}
\comment{
\FS{What is the relation between $Y$ and $V$? From the
text, both $Y = M V$ and $Y = V$ would be possible. Typically, in
settings like this, $Y$ would be dimensionless, meaning that $Y = V \:
t_0/\sigma$ or $Y= M V \: t_0/m \sigma$.
\MHB{Ich bin auch nicht mehr sicher: Niklas, kannst Du bitte Friederike und
mir kommunizieren, ob das die Groesse der Ableitung von Y oder von V ist.
Achtung: Man muss die Legende in Figure 4.1 ggf. anpassen.}
\NB{Es handelt sich um die Größe der Ableitung von $C_Y$, wobei $Y$ so normiert ist, dass $C_Y(0) = I$.}}
\MHB{Liebe Friederike, ich glaube, aus \req{interpol} folgt: $Y = \sqrt{k_BT/M}\,V$ und
$V$ hat Einheit $\sigma/t_0$? und $k_BT = m\sigma^2/t_0^2$, also $[Y] = \sqrt{m/M}$?}
}
\MHB{Accordingly, this artificial \FS{instantaneous damping} term is tolerable from a physical point of view.}
\MHB{Ich bin nicht sicher, was wir hier schreiben wollen; am besten, wir
diskutieren das in personam :-)}

\FS{FS: I also don't understand this statement. This
''artificial drift'' is just an instantaneous contribution to the
memory kernel, right?  It does not really introduce new physics, but
it should not be there from a mathematical point of view. To
understand whether it is tolerable or not, one would have to compare
$D$ with the integral $\int \tilde{\gamma}(s) \text{d}s$.}}

Concerning Method~B (from Section~\ref{Subsec:rat_app})
the AAA algorithm determines a rational function with seven admissible poles, 
from which one is positive and the remaining six come in three complex conjugate pairs.
The algorithm therefore proceeds with a Prony series Ansatz with $p=7$ terms,
and the computation of the corresponding matrix coefficients $\Gamma_j$
is carried out as described in Section~\ref{Subsec:rat_app}. 
The resulting system matrix $A$ -- constructed as in Appendix~\ref{Subsec:sysmat} --
has dimension $dp = 14$, which corresponds to $N=12$ auxiliary variables 
for this system. 
It turns out that the singular Lur'e system~\req{singLure} is solvable,
and the solution can be found as in Appendix~\ref{App:ZZZ}, because
the two conditions \req{BC} and \req{BAC} are indeed satisfied.
In other words, the associated Prony series is also a valid 
autocorrelation function.

%Note that just because of the five scalar constraints in \req{fprime} and
%the four additional constraints in \req{f0},
%any Prony series approximation $\varphi\approx C_V$ provided by Method~B
%need to have at least $m=3$ exponentials with $d^2m=12$ corresponding entries 
%(i.e., free parameters) of the coefficient matrices $\Gamma_j$;
%in fact, with only three exponentials the fit of the data points will usually 
%be poor, and therefore $m\geq 4$ is more reasonable. Accordingly the corresponding 
%Langevin system will have at least $N=d(m-1)\geq 6$ auxiliary variables.

\begin{figure}
    \centering
    \includegraphics[width=\linewidth]{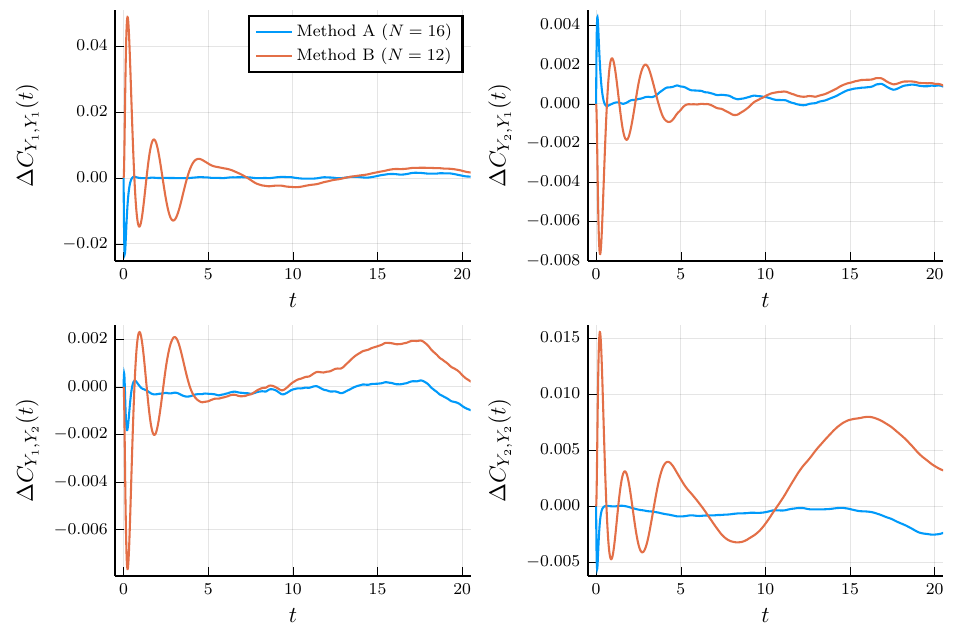}
    \caption{Differences $\Delta C_Y$ between the given velocity autocorrelation function 
    (same units) and the two approximations: 
    Method~A is the block Prony method from Section~\ref{Subsec:Prony} and 
    Method~B is the one from Section~\ref{Subsec:rat_app} based on 
    rational approximation. 
    $N$ is the number of auxiliary variables of the respective
    Langevin system. 
    Every panel shows the scalar graphs corresponding to the
    respective entry of the (matrix-valued) function.
    }
    \label{fig:example1-difference}
\end{figure}

Figure~\ref{fig:example1} displays the two constructed Prony series
approximations of the autocorrelation function, the one from Method~A in blue, 
and the one from Method~B in red.
Each subplot corresponds to one entry of the corresponding $2\times 2$ 
matrices. In these plots the blue line is hardly visible 
as it is almost covered by the red one. 
The inset plots therefore display enlarged versions of the corresponding graphs
restricted to large values of $t$.

Although not visible with the eye, the red line is -- by construction -- 
the graph of a function
with zero derivative at the origin (in each subplot), and it turns out that
it is for this reason that the most significant differences 
between the two approximations occur for times $t<5 \,t_0$. 
This can be seen in Figure~\ref{fig:example1-difference}, which displays 
the difference $\Delta C_Y$ between the two constructed Prony series 
$\varphi=C_Y$ and the input autocorrelation function $\frac{M}{k_BT}C_V$.
It also shows that the detection and proper reconstruction of the steep descent
of the autocorrelation function for $t < 1$, which is only encoded within the 
first few data points, poses a big challenge for both methods; accordingly,
the error of both methods is largest in this initial time interval.

As can be seen from the inset plots the approximations from both methods 
start to deviate from the given autocorrelation data far in the 
extrapolated regime, i.e., for times $t\geq 20 \,t_0$.
To be specific, both approximations converge to zero in this regime,
whereas the given data fail to do so because of noise. One may argue
that in this regime the computed approximations provide a better picture 
of the true autocorrelation functions than the given MD data.

%%%%%%%%%%%%%%%%%%%%%%%%%%%%%%%%%%%%%%%%%%%%%%%%%%%%%%%%%%%%
%\subsection{Example 5.2}
%\label{Subsec:Ex5.2}
%%%%%%%%%%%%%%%%%%%%%%%%%%%%%%%%%%%%%%%%%%%%%%%%%%%%%%%%%%%%
\begin{table}
    \centering
    \begin{tabular}{c | S[table-format=1.3] S[table-format=1.3] S[table-format=1.3] c | S[table-format=1.3] S[table-format=1.3] S[table-format=1.3] c}
        \toprule
        & \multicolumn{4}{c|}{block Prony method} & 
          \multicolumn{4}{c}{rational approx.\ method} \\[1ex]
        $n$ & {(a)} & {(b)} & {(c)} & $N$ & {(a)} & {(b)} & {(c)} & $N$ \\
        \midrule
        $15$ & 0.00130937 & 0.0241241 & 0.0241241 & 16 & 0.0186 & 0.0524 & 0.0524 & 12\\
        $16$ & 0.0010906 & 0.0224383 & 0.0224383 & 19 & 0.0347 & 0.0651 & 0.0651 & 12\\
        $17$ & 0.00195591 & 0.0222618 & 0.0222618 & 17 & 0.0397 & 0.0684 & 0.0684 & 12\\
        $18$ & 0.00246194 & 0.0216105 & 0.0216105 & 19 & 0.0494 & 0.0744 & 0.0744 & 12\\
        $19$ & 0.112403 & 0.11244 & 0.11244 & 18 & 0.0487 & 0.0740 & 0.0740 & 12\\
        $20$ & 0.015161 & 0.0177961 & 0.0177961 & 20 & 0.0163 & 0.0536 & 0.0536 & 16\\
        $21$ & 0.0143112 & 0.0200376 & 0.0200376 & 18 & 0.0249 & 0.0607 & 0.0607 & 16\\
        $22$ & 0.0331504 & 0.0332094 & 0.0332094 & 25 & 0.0159 & 0.0524 & 0.0524 & 16\\
        $23$ & 0.00687212 & 0.0224909 & 0.0224909 & 27 & 0.0146 & 0.0516 & 0.0516 & 16\\
        $24$ & 0.00347295 & 0.0215978 & 0.0215978 & 35 & 0.0147 & 0.0515 & 0.0515 & 16\\
        $25$ & 0.00159288 & 0.0217292 & 0.0217292 & 35 & 0.0157 & 0.0528 & 0.0528 & 16\\
        $26$ & 0.00205379 & 0.0176284 & 0.0176284 & 28 & 0.0158 & 0.0527 & 0.0527 & 16\\
        $27$ & 0.00212531 & 0.0211086 & 0.0211086 & 30 & 0.0184 & 0.0567 & 0.0567 & 16\\
        $28$ & 0.00392191 & 0.0204897 & 0.0204897 & 33 & 0.0184 & 0.0566 & 0.0566 & 16\\
        $29$ & 0.00183341 & 0.0206415 & 0.0206415 & 32 & 0.0183 & 0.0565 & 0.0565 & 16\\
        $30$ & 0.0012961 & 0.0200837 & 0.0200837 & 32 & 0.0183 & 0.0564 & 0.0564 & 16\\
        \bottomrule
    \end{tabular}
    \caption{Approximation errors 
    (same units as in Figure~\ref{fig:example1-difference}) 
    of both methods (a) at the given grid points,
    (b) in between the grid points, and (c) for the entire time interval $[0, 50]$,
    for varying number of input data. Also listed are the numbers $N$
    of respective auxiliary variables. 
    % Red entries in the left half mean that we had to remove additional eigenvalues (with negative real part) from the system matrix in order to achieve a valid autocorrelation function.
    }
    \label{tab:example2-interpolation-error}
\end{table}
Table~\ref{tab:example2-interpolation-error} compiles the results of further
test runs for the same example, but using more input data. 
More precisely, each run uses data from a different grid with the 
same grid spacing $\tau=0.5 \, t_0$, but now with $2n$ data points, 
where $n$ ranges up to 30. For each value of $n$ the table lists the maximum
approximation error at the respective grid points in column (a), 
the maximum approximation error in between these grid points in column (b), 
and the maximum approximation error up to $t=50 \, t_0$ 
in column (c) for each of the two methods. Each of these error numbers 
corresponds to the Frobenius norm of the difference $\Delta C_Y$ between the input 
autocorrelation function and its approximation $\varphi$ at a given time.
%For each $n=10, \ldots, 23$ it displays the maximum errors (Frobenius norm) of 
%the obtained VACF approximation compared to the input VACF
%\begin{enumerate}[label=(\alph*)]
%    \item of all points in the interpolation grid: 
%$t = 0, \tau, \ldots, (2n-1)\tau$
%    \item of all data points up to the last data point which belongs to the 
%interpolation grid: $t = 0, \tilde\tau, \ldots, (2n-1)\tau$
%where $\tilde\tau = 0.1$ \textcolor{red}{[unit]} is the size of the grid for 
%which the input data were simulated
%    \item of all data points: $t=0, \tilde\tau, \ldots, 599\tilde\tau = 59.9$ 
%\textcolor{red}{[unit]}
%\end{enumerate}

% In all test runs the second data point $Y_2$ had to be shifted downwards in Method~A to obtain a system matrix~\req{Langevin-nonzero} with an entry $-D$ for which $D+D^T$ is positive semidefinite. 
% In general these shifts range between $3\cdot10^{-3}$ and $6\cdot10^{-3}$ times the identity matrix; only in two cases the shifts were larger by a factor of up to five.
% Still, in two cases Method~A did not succeed to find a system matrix $A$ for which the resulting Prony series approximation $\varphi$ of the  autocorrelation function has a Fourier transform which is positive definite,  i.e., for which the Lur'e system has a solution.
% So no approximating Langevin system could be obtained in these two instances.

Method~A succeeds for every tested value of $n$, i.e.\ it yields
approximations of the given autocorrelation function, which are admissible
in the sense that they are autocorrelation functions of the $Y$-component
of the solution of an appropriate Langevin equation~\req{Langevin}.
The associated approximation errors are small in all cases except for the 
grid corresponding to $n=19$, where a pair of two complex conjugate spurious poles 
with comparably large coefficients 
\num[round-precision=4]{0.0044898727977808} has to be removed from the Prony
series, which results in a larger approximation error. 

Method~B also yields admissible approximations for all the data sets that 
we have tested. 
For the smaller grids, i.e., for $n$ up to 19, the amount of data has only 
a minor influence on the computed approximations in that the resulting
Prony series are very similar for these grids.
For $n=20$ and beyond, two further (complex conjugate) exponentials 
(and $2d=4$ further auxiliary variables) are chosen for a better match of the 
additional data points; this corresponds to one
additional step of the greedy iteration in the AAA method 
(see Appendix~\ref{Subsec:AAA}). For the two cases $n=20$ and $n=21$
this additional iteration was enforced manually, since the approximation 
chosen with the default parameters of our implementation would have been
too poor otherwise. For $n>21$ the algorithm made the additional iteration
by itself. 
%As can be seen in Figure~\ref{fig:example1}, 
%for $n\geq 20$ the data for the additional grid points pick up a lot of noise. 
%According to the corresponding error numbers in 
%Table~\ref{tab:example2-interpolation-error}
%the quality of the output of Method~B is not really affected by this
%because the method avoids an exact interpolation of the given data.

From Table~\ref{tab:example2-interpolation-error} it can further be seen
that the approximation error of Method~A is usually smaller
than the error of Method~B by a factor of two to three,
except when $n=19$, where the error in Method~A is exceptionally large. 
On the other hand, Method~A requires a larger number of auxiliary variables.

In summary, the numerical examples show that both methods have high potential 
to provide excellent approximations of vector-valued stochastic processes 
coming from GLEs. For our particular test case
Method~B shows a slightly better performance, as it manages with
fewer auxiliary variables, and at the same time fulfills the 
side constraints \eqref{eq:fprime} and \eqref{eq:f0}.
Method~A on the other hand yields better approximations errorwise, but in general needs up to twice the amount of auxiliary variables than Method~B.
% apart of that the two methods yield approximations with comparable approximation errors. 

%%%%%%%%%%%%%%%%%%%%%%%%%%%%%%%%%%%%%%%%%%%%%%%%%%%%%%%%%%%%
%\section{Discussion}
%\label{Sec:Discussion}
%%%%%%%%%%%%%%%%%%%%%%%%%%%%%%%%%%%%%%%%%%%%%%%%%%%%%%%%%%%%

%%%%%%%%%%%%%%%%%%%%%%%%%%%%%%%%%%%%%%%%%%%%%%%%%%%%%%%%%%%
\section{Conclusion}
\label{Sec:Conclusion}
%%%%%%%%%%%%%%%%%%%%%%%%%%%%%%%%%%%%%%%%%%%%%%%%%%%%%%%%%%%
We have presented an extension of a recently proposed framework
for constructing Markovian embeddings of GLEs
directly from trajectory data to multidimensional systems. 
The approach bypasses the intermediate step of reconstructing 
a full memory kernel and instead identifies a minimal Markovian 
representation whose autocorrelation function matches that of the 
observed dynamics.

Two numerical strategies have been developed and compared. Method
A uses a Prony series expansion, extending a scheme proposed in
Ref.~\cite{BSJSH21} for the scalar case to vector-valued stationary
solutions of GLEs. Method B introduces a 
generating function, following Ref.~\cite{Hanke24},
which is then approximated by a rational function using
the AAA algorithm. Application to a two-dimensional test system — 
a rigid S-shaped probe particle in a solvent of Langevin particles 
— demonstrates that both methods can accurately reproduce velocity 
autocorrelation functions obtained from molecular dynamics 
simulations.

Approximation errors of the resulting autocorrelation functions and 
the number of auxiliary variables have been documented for different sizes 
of the input data sets to demonstrate the performance and the potential of 
the two methods. For our test example the block version of Prony's method 
(Method~A of Section~\ref{Subsec:Prony}) yields smaller approximation 
errors than the method based on rational approximation 
(Method~B of Section~\ref{Subsec:rat_app}), 
whereas the latter avoids an artificial instantaneous damping term
and requires somewhat fewer auxiliary variables.
The comparative analysis thus suggests that Method A is preferable 
when accuracy is paramount, while Method B is advantageous when 
model parsimony is desired.

The framework is broadly applicable to the data-driven construction
of reduced stochastic models in higher dimensions. In this work, we
did not consider GLEs with additional 
deterministic forces, or other position-dependent phenomena. 
This is left for future work.

\section*{Acknowledgements}
%%%%%%%%%%%%%%%%%%%%%%%%%%%%%%%%%%%%%%%%%%%%%%%%%%%%%%%%%%%
The research leading to this work
was funded by the Deutsche Forschungsgemeinschaft (DFG, German Research Foundation) in the framework to the collaborative research center "Multiscale Simulation Methods for Soft-Matter Systems" (TRR 146, Subproject A3) under Project No.~233630050.

\section*{Data Availability}
The data of the velocity autocorrelation function and the implemented code 
for the methods described in 
Section~\ref{Subsec:Prony}/Appendix~\ref{App:Lanczos} and 
Section~\ref{Subsec:rat_app}/Appendix \ref{App:rat_app} 
are openly available on Github
({\tt https://github.com/bmaximi/vector-memory}).

\appendix
%%%%%%%%%%%%%%%%%%%%%%%%%%%%%%%%%%%%%%%%%%%%%%%%%%%%%%%%%%%
\section{The Block Lanczos approach}
\label{App:Lanczos}
%%%%%%%%%%%%%%%%%%%%%%%%%%%%%%%%%%%%%%%%%%%%%%%%%%%%%%%%%%%
\subsection{Outline of the Lanczos method} 
\label{Subsec:Lanczos}
%%%%%%%%%%%%%%%%%%%%%%%%%%%%%%%%%%%%%%%%%%%%%%%%%%%%%%%%%%%
The algorithm from \cite{Freund01} computes two finite sequences of polynomials
$\psi_k$ and $\phi_k$ of a real variable $x$ of degree $k$, respectively, 
with $k$ running from zero to $(n+1)d-1$, such that
\be{biorth}
	\Bigl|\scalarprod{\psi_k}{\phi_l}[\mathcal{H}]\Bigr| \,=\, \delta_{kl}\,,
\ee
where $\scalarprod{\placeholder}{\placeholder}[\mathcal{H}]$ is the bilinear
form introduced in \eqref{eq:lanczos-bilinear-form-matrix-product-shape}.

These polynomials are generated recursively by first initializing
\[
   \psi_k \,=\, \phi_k \,=\, x^k\,, \qquad k=0,\dots,d-1\,,
\]
and then, defining
\be{alphak}
   \alpha_{k,k+d}\,\psi_{k+d} \,=\,
   x^d\psi_k \,-\, \sum_{i=0}^{k+d-1} \alpha_{k,i}\,\psi_i 
\ee
and
\be{betak}
   \beta_{k,k+d}\,\phi_{k+d} \,=\,
   x^d\phi_k \,-\, \sum_{i=0}^{k+d-1} \beta_{k,i}\,\phi_i
\ee
for $k=0,1,2,\dots,nd-1$, with
\be{alphabeta}
   \alpha_{k,i} \,=\, \frac{\scalarprod{x^d\psi_k}{\phi_i}[\mathcal{H}]}{\scalarprod{\psi_i}{\phi_i}[\mathcal{H}]}
   \quad \text{and} \quad
   \beta_{k,i} \,=\, \frac{\scalarprod{\psi_i}{x^d\phi_k}[\mathcal{H}]}{\scalarprod{\psi_i}{\phi_i}[\mathcal{H}]}
\ee
for $i=0,\dots,k+d-1$. 
We choose the positive scaling factors $\alpha_{k,k+d}$ and $\beta_{k,k+d}$
to have the same value, and this value is chosen such that
\[
   \Bigl|\scalarprod{\psi_{k+d}}{\phi_{k+d}}[\mathcal{H}]\Bigr| \,=\, 1\,.
\]
This is the point where the algorithm can fail, namely, 
if no such normalization is possible;
but as mentioned before, we have never encountered a failure in our
numerical tests.
Take note that the evaluation of the bilinear form requires the coefficients
of the respective polynomials, when expanded in powers of $x$; 
compare \req{lanczos-bilinear-form-matrix-product-shape}. 

Due to the block Hankel structure of the matrix $\mathcal{H}$ 
the bilinear form satisfies
\[
   \scalarprod{x^d\psi}{\phi}[\mathcal{H}]
   \,=\, \scalarprod{\psi}{x^d\phi}[\mathcal{H}]
\]
for any two polynomials $\phi$ and $\psi$ of degree $nd-1$ or less. 
It therefore follows from \req{biorth} that
\[
   \alpha_{k,i} = \beta_{k,i} = 0 \qquad \text{for $0\leq i<k-d$}\,,
\]
and hence, the recursion coefficients $\alpha_{k,i}$ make up the entries of
a real $nd\times nd$ matrix 
\be{T}
%   T \,=\, \begin{cmatrix}
%              \alpha_{0,0} & \cdots &\alpha_{0,d} & \!\!\!\!\!0 
%                 & \cdots & 0\\[1ex]
%              \vdots & \ddots & & &%\!\!\!\!\!\ddots & %\!\!\!\!\!\ddots
%                 & \vdots \\[1ex]
%              \alpha_{d,0} &   & \alpha_{d,d} & \!\!\!\!\!\alpha_{d,d+1}
%                 & \cdots & 0\\[1ex]
%              0 & \ddots & & \!\!\!\!\!\ddots & &
%                  \alpha_{(n-1)d-1,nd-1}\\[1ex]
%              \vdots & \ddots & \ddots & & \ddots & \vdots  \\[1ex]
%              0 & \cdots & 0 & \!\!\!\!\!\alpha_{nd-1,(n-1)d-1} & \cdots 
%                 & \alpha_{nd-1,nd-1}
%           \end{cmatrix}
%   \in \R^{nd\times nd}.
   T \,=\, \begin{cmatrix}
              \alpha_{0,0} & \cdots &\alpha_{0,d} & \!\!\!\!\!0 
                 & \cdots & 0\\[1ex]
              \vdots & & & \ddots &%\!\!\!\!\!\ddots & %\!\!\!\!\!\ddots
                 & \vdots \\[1ex]
              \alpha_{d,0} & & & & \ddots & 0\\[1ex]
              0 & \ddots & & &  & \alpha_{(n-1)d-1,nd-1}\\[1ex]
              \vdots & & \ddots & &  & \vdots  \\[1ex]
              0 & \cdots & 0 & \alpha_{nd-1,(n-1)d-1} & \cdots 
                 & \alpha_{nd-1,nd-1}
           \end{cmatrix}
\ee
of upper and lower bandwidth $d$.
This is the matrix $T$ which satisfies \req{J-interpolation-property};
compare Freund~\cite{Freund95} for a proof.
Take note that the coefficients $\alpha_{k,i}$ with $i \geq nd$ for the
last block of indices $k\geq (n-1)d$ are irrelevant for the definition 
of this matrix.

%%%%%%%%%%%%%%%%%%%%%%%%%%%%%%%%%%%%%%%%%%%%%%%%%%%%%%%%%%%
\subsection{Spurious eigenvalues of $\boldsymbol{T}$}
\label{Subsec:Lanczos-eigenvalues}
%%%%%%%%%%%%%%%%%%%%%%%%%%%%%%%%%%%%%%%%%%%%%%%%%%%%%%%%%%%
The following spectral modifications of $T$ are similar to \cite{BSJSH21}. 
Throughout we assume that $T$ is diagonalizable.

To remove spurious exponentials from the Prony series, we consider the 
eigenvalue decompositions
\begin{equation*}
	T = X\Lambda_TX^{-1} \qquad\text{and}\qquad A = X\Lambda_A X^{-1}
\end{equation*}
of $T$ and $A$ with $\Lambda_T = \diag(z_1, \ldots, z_{nd})$ and 
$\Lambda_A = \diag(\lambda_1, \ldots, \lambda_{nd})$, 
where $z_k=e^{\tau\lambda_k}$.
%$\lambda_k = \frac{1}{\tau}\log z_k$.
Let $q\in\N$ be such that $\lambda_{q+1}, \ldots, \lambda_{nd}$ be all 
the eigenvalues of $A$ with positive real part. 
Then we replace $\Lambda_A$ by 
$\widetilde\Lambda_A \defin \diag(\lambda_1, \ldots, \lambda_q)$ 
and $X$ by a matrix $\widetilde X$ which is obtained by removing the last 
$nd-q$ columns (corresponding to eigenvectors for the spurious
exponents) and certain rows from $X$. 
In principle any $nd-q$ rows can be removed from $X$
as long as the first $d$ rows are kept and the resulting matrix $\widetilde X$ 
is nonsingular. In the scalar case considered in \cite{BSJSH21} we always 
removed the last $n-q$ rows, but for $d>1$ this choice can fail.   
To enhance stability we construct $\widetilde X$ by copying the first $d$ rows
of $X$, and then appending one by one the particular row of $X$ to
$\widetilde X$, for which the norm of its orthogonal complement to the span of 
the already chosen rows divided by its norm is largest. 
\section{The rational approximation approach}
\label{App:rat_app}
%%%%%%%%%%%%%%%%%%%%%%%%%%%%%%%%%%%%%%%%%%%%%%%%%%%%%%%%%%%
\subsection{The AAA algorithm}
\label{Subsec:AAA}
%%%%%%%%%%%%%%%%%%%%%%%%%%%%%%%%%%%%%%%%%%%%%%%%%%%%%%%%%%%
We choose the AAA algorithm~\cite{NST18} to solve the rational approximation 
problem. More precisely, since the function $F$ of \req{gen-Func} 
to be approximated is
matrix-valued, we use a matrix-valued variant of the AAA algorithm 
from Gosea and G\"uttel~\cite{GoGue21}, which goes back to 
Lietaert et al.~\cite{LMPV22}. 

We let $\ZZ=\{\zeta_l\}$ be the elements of an equiangular grid with
$\pmax$ grid points on a circle $|\zeta|=\rho>1$ in the complex plane,
which are symmetric with respect to the real axis and include the two
real points $\pm\rho$, 
and approximate the associated function values $F(\zeta_l)$ by 
\[
   F_l \,=\, \sum_{\nu=0}^{2n-1} Y_\nu \zeta_l^{-\nu-1}\,.
\]
Below, the matrix entries of $F_l$ will be denoted by $F^l_{i,j}$, 
$i,j=1,\dots,d$. Concerning the particular choice of the grid,
we found in our numerical experiments that $\pmax=100$ and 
\bdm
   \rho\,=\,\bigl(1/\delta\bigr)^{\frac{1}{2n}}
\edm
with $\delta=10^{-6}$ is an appropriate choice for this particular
example. %Here $\delta$ serves as an estimate of the noise level in the data. 
These parameters should be adapted when other data sets are considered.

%The method assumes that we are given a finite set of $M$ support points 
%$\ZZ=\{z_k\}\subset \C$ with associated function values $F(z_k)$. 
The AAA Algorithm determines a rational function in barycentric form
\be{Bary}
    r(z) = \sum_{k=1}^p \frac{w_kF_k}{z - \zeta_k} \Big/ 
           \sum_{k=1}^p \frac{w_k}{z - \zeta_k}\,,
\ee
where $\zeta_k\in \ZZ$, $k=1,\dots,p<\pmax$, are the so-called support points, 
and $w_k$, $k=1,\ldots,p$, are suitable (complex) scalar weights.
Under the assumption that these weights are nonzero, the function~\req{Bary}
interpolates the given values $F_k$ at all the support points, i.e.,
\[
   r(\zeta_k) \,=\, F_k\,, \qquad k=1,\dots,p\,.
\]
%The term $w_0$ in the denominator, which does not appear in \cite{GoGue21,NST18}, 
%has been introduced to cope for the additional interpolation constraint 
%$r(\infty) = 0$, which naturally occurs in our application, 
%compare~\req{rat-app}. 
We take care (see below) 
%that $w_0\in\R$ and 
that the weights associated 
with complex conjugate support points are also complex conjugate, so that 
$r(\overline{z})=\overline{r(z)}$ for every $z\in\C$, and hence, $r$ has 
a representation as a quotient of two polynomials with real coefficients
(which albeit is never explicitly computed).
%Note that every entry $r_{ij}(z)$ of 
%$r(z)$ is a scalar rational function in barycentric form, and 

The algorithm employs a greedy iterative scheme. 
%Starting with $m=0$, i.e., with $r=0$ in \req{Bary}, 
In each iteration a new support point $\zeta_l\in \ZZ$ is chosen, for which the
Frobenius norm of the current nonlinear residual $\|F_l-r(\zeta_l)\|_F$ is 
maximal.
%from the subset of support points that have not yet been chosen, where the nonlinear
%residual $\Vert F(z) - r(z) \Vert_F$ of the previous step takes its maximum value.
We also include the complex conjugate point 
$\zeta_{l'}=\overline{\zeta}_l$ to the list of support points; 
%since $F_l$ and $F_{l'}$ are complex conjugate matrices and 
%$r(\zeta_{l'})=\overline{r(\zeta_l)}$, 
the two associated nonlinear residuals share the same Frobenius norm. 
Then the weights $w_1,\ldots,w_p$ for \req{Bary} are chosen such that they 
solve a linearized least-squares problem over the remaining points in $\ZZ$:
with $\pi$ denoting the sum in the denominator in \req{Bary}, we minimize
\[
    \sum_{l=p+1}^\pmax 
    \Bigl\| \pi(\zeta_l) \bigl(F_l - r(\zeta_l)\bigr) \Bigr\|_F^2 
    = \sum_{l=p+1}^\pmax \left\Vert 
       \sum_{k=1}^p w_k\,\frac{F_l - F_k}{\zeta_l - \zeta_k}\right\Vert_F^2 
       \to \textrm{min.}\,,
\]
subject to the constraint that the weight vector $w=[w_1,\dots,w_p]^T$ has
unit (Euclidean) norm. We can write this in the form
\be{aaa-LSQ}
    \Vert L w\Vert_2 \to \textrm{min.}\,, \qquad \Vert w \Vert_2 = 1\,,
\ee
with a Loewner matrix
%\[
%    L = \begin{cmatrix}
%        F^{m+1}_{1,1} & \frac{F^{m+1}_{1,1} - F^{1}_{1,1}}{\zeta_{m+1} - \zeta_1} & \dots & \frac{F^{m+1}_{1,1} - F^m_{1,1}}{\zeta_{m+1} - \zeta_m}\\
%        \vdots & \vdots &  & \vdots\\
%        F^M_{1,1} & \frac{F^M_{1,1} - F^1_{1,1}}{\zeta_{M} - \zeta_1} & \ldots & \frac{F^M_{1,1} - F^m_{1,1}}{\zeta_{M} - \zeta_m}\\[1ex]
%        F^{m+1}_{2,1} & \frac{F^{m+1}_{2,1} - F^1_{2,1}}{\zeta_{m+1} - \zeta_1} & \dots & \frac{F^{m+1}_{2,1} - F^m_{2,1}}{\zeta_{m+1} - \zeta_m}\\
%        \vdots & \vdots &  & \vdots\\
%        F^M_{2,1} & \frac{F^M_{2,1} - F^1_{2,1}}{\zeta_{M} - \zeta_1} & \ldots & \frac{F^M_{2,1} - F^m_{2,1}}{\zeta_{M} - \zeta_m}\\
%        \vdots & \vdots &  & \vdots\\
%        F^{m+1}_{d,d} & \frac{F^{m+1}_{d,d} - F^1_{d,d}}{\zeta_{m+1} - \zeta_1} & \dots & \frac{F^{m+1}_{d,d} - F^m_{d,d}}{\zeta_{m+1} - \zeta_m}\\
%        \vdots & \vdots & & \vdots\\
%        F^M_{d,d} & \frac{F^M_{d,d} - F^1_{d,d}}{\zeta_{M} - \zeta_1} & \ldots & \frac{F^M_{d,d} - F^m_{d,d}}{\zeta_{M} - \zeta_m}\\
%    \end{cmatrix} .
%\]
\[
    L = \begin{cmatrix}
        \frac{F^{p+1}_{1,1} - F^{1}_{1,1}}{\zeta_{p+1} - \zeta_1} & \dots & \frac{F^{p+1}_{1,1} - F^p_{1,1}}{\zeta_{p+1} - \zeta_p}\\
        \vdots & & \vdots\\
        \frac{F^\pmax_{1,1} - F^1_{1,1}}{\zeta_\pmax - \zeta_1} & \ldots & \frac{F^\pmax_{1,1} - F^p_{1,1}}{\zeta_\pmax - \zeta_p}\\[1ex]
        \frac{F^{p+1}_{2,1} - F^1_{2,1}}{\zeta_{p+1} - \zeta_1} & \dots & \frac{F^{p+1}_{2,1} - F^p_{2,1}}{\zeta_{p+1} - \zeta_p}\\
        \vdots & & \vdots\\
        \frac{F^\pmax_{2,1} - F^1_{2,1}}{\zeta_{\pmax} - \zeta_1} & \ldots & \frac{F^\pmax_{2,1} - F^p_{2,1}}{\zeta_{\pmax} - \zeta_p}\\
        \vdots & & \vdots\\
        \frac{F^{p+1}_{d,d} - F^1_{d,d}}{\zeta_{p+1} - \zeta_1} & \dots & \frac{F^{p+1}_{d,d} - F^p_{d,d}}{\zeta_{p+1} - \zeta_p}\\
        \vdots & & \vdots\\
        \frac{F^\pmax_{d,d} - F^1_{d,d}}{\zeta_\pmax - \zeta_1} & \ldots & \frac{F^\pmax_{d,d} - F^p_{d,d}}{\zeta_\pmax - \zeta_p}\\
    \end{cmatrix} .
\]
This means that any solution of \req{aaa-LSQ} is a unit eigenvector of the 
smallest eigenvalue of $L^\ast L$, and one of them (at least) has complex 
conjugate entries associated with any pair of complex conjugate support points. 
To determine one of those, one can take any nonzero vector $\widetilde{w}$ 
from this eigenspace and set
\[
    w = \left\{
        \begin{array}{ll}
           \rmi\,\widetilde{w}/\|\widetilde{w}\|_2\, &\quad \text{if}\
           \widetilde{w} = -\overline{P\widetilde{w}}\,, \\[1ex]
           (\widetilde{w} + \overline{P\widetilde{w}})/
           \bigl\|\widetilde{w} + 
                  \overline{P\widetilde{w}}\hspace*{0.1ex}\bigr\|_2\, &\quad
           \text{if}\
           \widetilde{w} \neq -\overline{P\widetilde{w}}\,, 
        \end{array}\right.
\]
where $P$ denotes the permutation matrix that swaps the entries corresponding 
to complex conjugate support points. It is not difficult to see that
this vector $w$ is an element from the very same eigenspace with the 
desired property. 

%The rational approximation $r^{(m)}(z)$ can then be readily computed by using 
%the identity \req{Bary} to evaluate the nonlinear residuals. 
In our implementation we initialize the greedy iterative scheme with $p=2$ 
support points $\zeta_1=\rho$ and $\zeta_2=-\rho$, and we terminate the iteration 
when all nonlinear residuals $\|F_k-r(\zeta_k)\|_F$ are below a tolerance of $10^{-4}$ in norm and the rational approximation has at least 4 admissible poles, since the algorithm described in Section~\ref{Subsec:AAA} requires
the poles of the approximation~\req{Bary}. These are given as the $p-1$ finite
eigenvalues $z$ of the $(p+1)\times(p+1)$ generalized eigenvalue problem
\[
    \begin{cmatrix} 
       0 & w_1 & w_2 & \ldots & w_m\\ 
       1 & \zeta_1 &&&\\ 1 && \zeta_2 && \\ \vdots &&& \ddots & \\ 1 &&&& \zeta_m
    \end{cmatrix} 
    \,=\,
    z \begin{cmatrix}
        \,0 & \phantom{w_1} & \phantom{w_2} && \\ 
        & 1 &&& \\ && 1 && \\ &&& \ddots & \\ &&&& 1\\
      \end{cmatrix} \,.
\]
Take note that two further generalized eigenvalues of this problem
are infinite \emph{per se}.
%and the remaining $m$ are the poles of the rational approximation. 

%%%%%%%%%%%%%%%%%%%%%%%%%%%%%%%%%%%%%%%%%%%%%%%%%%%%%%%%%%%
\subsection{Construction of the system matrix}
\label{Subsec:sysmat}
%%%%%%%%%%%%%%%%%%%%%%%%%%%%%%%%%%%%%%%%%%%%%%%%%%%%%%%%%%%
In the following we present one possibility to assemble the system matrix $A$
for a given set of exponents $\lambda_j$ and associated coefficient matrices 
$\Gamma_j$ of the real-valued Prony series $\varphi$ of \req{Prony}. 
We start with the real Jordan canonical matrix
\be{J}
%    J = I \otimes \begin{cmatrix}
%        \lambda_1 \\ & \lambda_2 \\ && \ddots\\
%        &&& \lambda_r \\
%        &&&& \left(\begin{array}{cc}
%            \mathrm{Re}\,\lambda_{r+1} & \mathrm{Im}\,\lambda_{r+1}\\
%         -\mathrm{Im}\,\lambda_{r+1} & \mathrm{Re}\,\lambda_{r+1}\\
%        \end{array} \right) \\
%        &&&&&& \ddots\,
%    \end{cmatrix}\,,
    J = I \otimes \begin{cmatrix}
        \lambda_1 \\ & \lambda_2 \\ && \ddots\\
        &&& \lambda_q \\
        &&&& \phantom{-}\mathrm{Re}\,\lambda_{q+1} & 
             \mathrm{Im}\,\lambda_{q+1}\\
        &&&& -\mathrm{Im}\,\lambda_{q+1} & \mathrm{Re}\,\lambda_{q+1}\\
        &&&&&& \ddots\,
    \end{cmatrix}\,,
\ee
written as a Kronecker tensor product with a $d\times d$ identity matrix $I$.
The leading part of the second factor in \req{J} consists of the $q$, say, 
real-valued exponents, 
followed by $2\times 2$ rotation matrices for the complex conjugate pairs, 
where $\lambda_{q+2} = \overline{\lambda}_{q+1}$, and so on. 
%In \req{J}, the leading factor $I$ in the Kronecker tensor product is the
%$d\times d$ identity matrix.
Furthermore let
\be{svd-coeff}
    \Gamma_j = \sum_{i=1}^d \sigma_{(j-1)d+i}\,u_{(j-1)d+i}\,v_{(j-1)d+i}^* 
\ee
be the singular value decompositions of the coefficient matrices
$\Gamma_j$, $j=1,\dots,p$. Then the real-valued matrices
\[
    U \,=\, 
    \begin{cmatrix}
        \sqrt{\sigma_1}\,u_1 & \dots & \sqrt{\sigma_{dq}}\,u_{dq} &
        \sqrt{2\sigma_{dq+1}}\ \mathrm{Re}\, u_{dq+1} & 
        \sqrt{2\sigma_{dq+1}}\ \mathrm{Im}\, u_{dq+1} & \dots \phantom{x}
    \end{cmatrix}^T\!\phantom{,}
\]
and
\[
    V \,=\,
    \begin{cmatrix}
        \sqrt{\sigma_1}\,v_1 & \dots & \sqrt{\sigma_{dq}}\,v_{dq} &
        \sqrt{2\sigma_{dq+1}}\ \mathrm{Re}\, v_{dq+1} &
        -\sqrt{2\sigma_{dq+1}}\ \mathrm{Im}\, v_{dq+1} & \dots \phantom{x}
    \end{cmatrix}^T\!,
\]
defined in accordance with \req{J}, satisfy
\be{f-UJV}
    \varphi(t) = U^Te^{tJ}V \,, \qquad t\ge 0 \,.
\ee

Furthermore, since $U^TV = \varphi(0) = I$ by virtue of \req{Prony} and
\req{LSQ}, there exist nonsingular real-valued matrices $X$ with
\be{XV-XU}
    XV = E \quad \mbox{and} \quad X^T E = U\,,
\ee
and hence, the desired representation~\req{desire} is achieved for
\be{A-construction}
    A \,=\, X J X^{-1}.  
\ee
A construction of such an appropriate basis transformation matrix $X$ 
is presented in Appendix~\ref{App:Lemma}.
%of \req{XV-XU} let
%\[
%V^T = Q \begin{cmatrix} R_0\\ 0\\ \end{cmatrix}
%\]
%be the QR-decomposition of $V^T$, where $Q\in\R^{dm\times dm}$ is an 
%orthogonal matrix, and the nonsingular block $R_0\in\R^{d\times d}$ is upper 
%triangular. Extending $R_0$ by an identity matrix block $I_0$ to the 
%$dm\times dm$ block matrix
%\[
%    R = \begin{cmatrix}
%        R_0 & 0\\ 0 & I_0\\
%    \end{cmatrix}\,,
%\]
%and defining
%\be{X}
%    X = R^{-1}Q^T + E \bigl(U^T - V^T(VV^T)^{-1}\bigr)^T \,,
%\ee
%it is not difficult to see that $X$ is nonsingular and satisfies \req{XV-XU}.

Note that the final system matrix has dimension $dp\times dp$, unless
some of the singular values $\sigma_i$ in \req{svd-coeff} happen to be zero.
In such a rare case one can simply eliminate the corresponding rows from $U$ 
and $V$
and the associated diagonal entries (or $2\times 2$ block entries) from $J$.

%%%%%%%%%%%%%%%%%%%%%%%%%%%%%%%%%%%%%%%%%%%%%%%%%%%%%%%%%%%
\section{Solving the Lur'e equations numerically}
\label{App:ZZZ}
%%%%%%%%%%%%%%%%%%%%%%%%%%%%%%%%%%%%%%%%%%%%%%%%%%%%%%%%%%%
In the regular case, where $R = D+D^T$ is positive definite, solving 
\req{Lure} amounts to solving the algebraic Riccati equation
\be{Riccati}
    P\Sigma_0 + \Sigma_0 P^T + \Sigma_0 BR^{-1}B^T\Sigma_0 + CR^{-1}C^T = 0 
\ee
for (the symmetric matrix) $\Sigma_0$, where we have set
\[
   P \,=\, A_0-CR^{-1}B^T
\]
for brevity. We refer to Bini et al~\cite{BIM12} for an overview of
possible numerical algorithms for solving~\req{Riccati};
in our code we have used the algorithm suggested by Laub~\cite{Laub}.
The left-hand side of \req{Riccati} can be rewritten as
\[
   A_0\Sigma_0+\Sigma_0A_0^T \,+\, (C-\Sigma_0B)R^{-1}(C-\Sigma_0B)^T\,.
\]
Therefore, if we take $K$ to be the Cholesky factor of $R$ and let
\[
   L \,=\, (C-\Sigma_0B)K^{-T}\,,
\]
then it follows that \req{Riccati} is equivalent to
\[
   A_0\Sigma_0 + \Sigma_0 A_0^T \,=\, -LL^T\,.
\]
Accordingly, the (regular) Lur'e equations~\req{Lure} are satisfied with
this choice of $\Sigma_0$, $K$, and $L$.

In the singular case, where $D=0$, we let
\[
   S \,=\, B^TC \,\in\, \R^{d\times d}\,,
\]
which we assume to be symmetric and positive definite; compare~\req{BC}.
As exemplified in Appendix~\ref{App:Lemma} there exists a nonsingular 
transformation matrix 
$X\in\R^{N\times N}$, such that
\be{XLure}
    XC = E_0S^{1/2} \quad \mbox{and} \quad 
    X^T E_0 = BS^{-1/2} \,,
\ee
where $E_0\in\R^{N\times d}$ is defined like $E$ in \req{E}. 
%\[
%   D_1 + D_1^T \,=\, K_1K_1^T\,, \quad
%   A_1\Sigma_1+\Sigma_1A_1^T \,=\, -L_1L_1^T\,,\quad
%   C_1 - \Sigma_1 B_1 \,=\, L_1K_1^T\,,
%\]
Then it follows from \req{XLure} that the solution $\Sigma_0$ 
of \req{singLure} satisfies
\[
   (X\Sigma_0 X^T)E_0 \,=\, X\Sigma_0 BS^{-1/2} \,=\, XCS^{-1/2} \,=\, E_0\,,
\]
i.e., that
\be{XSigma0X}
   X\Sigma_0 X^T \,=\, \begin{cmatrix} I & 0\\ 0 & \Sigma_1 \end{cmatrix}
\ee
for some sysmmetric positive definite $(N-d)\times(N-d)$ matrix block
$\Sigma_1$. We further conclude from \req{singLure} that
\be{Lyapunov0}
\begin{aligned}
   &(XA_0 X^{-1})(X\Sigma_0 X^T) \,+\, (X\Sigma_0 X^T)(XA_0 X^{-1})^T\\
   &\qquad \qquad \,=\, X(A_0\Sigma_0+\Sigma_0A_0^T)X^T \,=\, -(XL)(XL)^T\,.
\end{aligned}
\ee
Decomposing the two matrices
\be{XL}
   XL \,=\, \begin{cmatrix} K_1 \\ L_1 \end{cmatrix} \qquad \text{and} \qquad
   XA_0 X^{-1} \,=\, \begin{cmatrix} -D_1 & B_1^T \\ -C_1 & A_1 \end{cmatrix}
\ee
accordingly in $d$ and $N-d$ rows and columns, it follows from
\req{XSigma0X} and \req{Lyapunov0} that
\be{Lure1}
   D_1 + D_1^T \,=\, K_1K_1^T\,, \quad
   A_1\Sigma_1+\Sigma_1A_1^T \,=\, -L_1L_1^T\,,\quad
   C_1 - \Sigma_1 B_1 \,=\, L_1K_1^T\,,
\ee
i.e., that $\Sigma_1$, $K_1$, and $L_1$ solve a \emph{regular} Lur'e system.
Since
\[
   D_1 \,=\, -E_0^T(XA_0X^{-1})E_0 
   \,=\, -(E_0^TX)A_0(X^{-1}E_0)
   \,=\, -S^{-1/2}B^TA_0CS^{-1/2}
\]
by virtue of \req{XLure}, $D_1+D_1^T$ is
positive definite when \req{BAC} is assumed to hold. 
Therefore, under the given assumptions the regular Lur'e system~\req{Lure1}
can be solved via an algebraic Riccati equation as described above, and
$\Sigma_0$ and $L$ are then given by \req{XSigma0X} and \req{XL}.

%It can be shown (compare~\cite{Hanke24})
%that the Positive Real Lemma applies to this reduced $N\times N$ system 
%matrix, meaning that the (regular) Lur'e equations
%\[
%   D_1 + D_1^T \,=\, K_1K_1^T\,, \quad
%   A_1\Sigma_1+\Sigma_1A_1^T \,=\, -L_1L_1^T\,,\quad
%   C_1 - \Sigma_1 B_1 \,=\, L_1K_1^T\,,
%\]
%have a solution $\Sigma_1\in\R^{(N-d)\times(N-d)}$, $L_1\in\R^{(N-d)\times d}$,
%and $K_1\in\R^{d\times d}$, with $\Sigma_1$ being symmetric and positive
%semidefinite. Since $D_1+D_1^T$ is assumed positive definite the solution
%can be obtained as above by solving a corresponding 
%Riccati equation~\req{Riccati}. 
%
%Now we let
%\[
%   \Sigma_0 \,=\, 
%   X^{-1}\begin{cmatrix} I & 0 \\ 0 & \Sigma_1 \end{cmatrix}X^{-T}
%   \qquad \text{and} \qquad 
%   L \,=\, X^{-1} \begin{cmatrix} K_1 \\ L_1 \end{cmatrix}.
%\]
%Then one can readily check that
%\[
%   A_0\Sigma_0 \,+\, \Sigma_0A_0^T \,=\, -LL^T\,.
%\]
%Furthermore, it follows from \req{XLure} that
%\[
%   \Sigma_0 B 
%   \,=\, X^{-1}\begin{cmatrix} I & 0 \\ 0 & \Sigma_1 \end{cmatrix} X^{-T}B
%   \,=\, X^{-1} E_0 S^{1/2} \,=\, C \,.
%\]
%In other words, $\Sigma_0$ and $L$ solve the 
%singular Lur'e equations~\req{singLure}.

%%%%%%%%%%%%%%%%%%%%%%%%%%%%%%%%%%%%%%%%%%%%%%%%%%%%%%%%%%%
\section{A useful basis transformation}
\label{App:Lemma}
%%%%%%%%%%%%%%%%%%%%%%%%%%%%%%%%%%%%%%%%%%%%%%%%%%%%%%%%%%%
For any $q>d$ let $E_q\in\R^{q\times d}$ be defined accordingly to $E$ in \req{E}.
Furthermore, let $U,V\in\R^{q\times d}$ be such that
\be{S}
   S \,=\, U^TV
\ee
is symmetric positive definite. Choose a matrix $Q\in\R^{q\times(q-d)}$
whose columns form a basis of the orthogonal complement of the range of $V$; 
such a basis can be obtained, for example, via a singular value decomposition
of $V$.
Then the matrix
\be{X-app}
   X \,=\, \begin{cmatrix} S^{-1/2}U^T \\ Q^T \end{cmatrix} 
   \,\in\, \R^{q\times q}
\ee
is nonsingular, for if $Xv=0$ for some $v\in\R^q$ then $Q^Tv=0$, i.e.,
$v=Vy$ for some $y\in\R^d$, and at the same time
\[
   0 \,=\, S^{-1/2} U^Tv \,=\, S^{-1/2}U^TVy \,=\, S^{1/2} y
\]
by virtue of \req{S}. Since $S$ is assumed to be positive definite, this
implies that $y=0$, and hence, $v=0$.

Using \req{X-app} and \req{S} we immediately see that the matrix $X$
provides a coordinate transformation with the two properties
\begin{subequations}
\label{X-properties}
\be{X1}
   X V \,=\, \begin{cmatrix} S^{-1/2}U^TV \\ 0 \end{cmatrix}
   \,=\, E_qS^{1/2}
\ee
and
\be{X2}
   X^TE_q \,=\, \begin{cmatrix} US^{-1/2} & Q \end{cmatrix} E_q
   \,=\, US^{-1/2}\,.
\ee
\end{subequations}

\bibliographystyle{siam}
\bibliography{references}

@book{AV73,
Author = {Anderson, B. D. O. and Vongpanitlerd, S.},
Title = {Network Analysis and Synthesis: A Modern Systems Theory
Approach}, 
Year = {1973},
Publisher = {Prentice-Hall},
Address = {Englewood-Cliffs, NJ},
}

@book{BIM12,
Author = {Bini, D. A. and Iannazzo, B. and Meini B.},
Title = {Numerical {S}olution of {A}lgebraic {R}iccati {E}quations},
Publisher = {SIAM},
Year = {2012},
Address = {Philadelphia},
}

@book{Bjor24,
Author = {Bj\"orck, \AA{}},
Title = {Numerical {M}ethods for {L}east {S}quares {P}roblems},
Publisher = {SIAM},
Year = {2024},
Address = {Philadelphia},
}

@article{BSJSH21,
Author = {Bockius, N. and Shea, J. and Jung, G. and Schmid, F. and Hanke, M.},
Title = {Model reduction techniques for the computation of extended Markov
   parameterizations for generalized Langevin equations},
Journal = {J. Phys.: Condens. Matter},
Year = {2021},
Volume = {33},
Number = {21},
Month = {MAY 26},
DOI = {10.1088/1361-648X/abe6df},
Article-Number = {214003},
Pages = {214003},
ISSN = {0953-8984},
EISSN = {1361-648X},
}

@article{CVR14,
Author = {Carof, Antoine and Vuilleumier, Rodolphe and Rotenberg, Benjamin},
Title = {Two algorithms to compute projected correlation functions in molecular
   dynamics simulations},
Journal = {J. Chem. Phys.},
Year = {2014},
Volume = {140},
Number = {12},
Month = {MAR 28},
DOI = {10.1063/1.4868653},
Article-Number = {124103},
Pages = {124103},
ISSN = {0021-9606},
EISSN = {1089-7690},
Unique-ID = {WOS:000334169000008},
}

@article{DBP09,
   title = {Langevin Equation with Colored Noise for 
       Constant-Temperature Molecular Dynamics Simulations},
   author = {Ceriotti, Michele and Bussi, Giovanni and Parrinello, Michele},
   journal = {Phys. Rev. Lett.},
   volume = {102},
   issue = {2},
   pages = {020601},
   numpages = {4},
   year = {2009},
   month = {Jan},
   publisher = {American Physical Society},
   doi = {10.1103/PhysRevLett.102.020601},
   url = {https://link.aps.org/doi/10.1103/PhysRevLett.102.020601}
 }

@article{CBP10,
     doi = {10.1021/ct900563s},
     year = 2010,
     month = {mar},
     publisher = {American Chemical Society ({ACS})},
     volume = {6},
     number = {4},
     pages = {1170-1180},
     author = {Michele Ceriotti and Giovanni Bussi and Michele Parrinello},
     title = {Colored-Noise Thermostats {\`{a}} la Carte},
     journal = {J. Chem. Theory Comp.}
 }

@article{CLL14,
 author = {Chen, Minxin  and Li, Xiantao  and Liu, Chun },
 title = {Computation of the memory functions in the generalized 
     Langevin models for collective dynamics of macromolecules},
 journal = {J. Chem. Phys.},
 volume = {141},
 number = {6},
 pages = {064112},
 year = {2014},
 doi = {10.1063/1.4892412},
 }

@article{FP79,
 author = {Ferrario, Mauro and Grigolini, P.},
 title = {{A generalization of the Kubo—Freed relaxation theory}},
 journal = {Chem. Phys. Lett.},
 volume = {62},
 number = {1},
 pages = {100 - 106},
 year = {1979},
 issn = {0009-2614},
 doi = {10.1016/0009-2614(79)80421-2},
 }

@inproceedings{Freund95,
Author = {Freund, R. W.},
Title = {{Computation of matrix Pad\'{e} approximations of transfer functions via a Lanczos-type process}},
Booktitle = {{Approximation Theory VIII, Vol. I}},
Editor = {C. K. Chi and L. L. Schumaker},
Publisher = {Birkh\"auser},
Address = {Boston},
Pages = {215 - 222},
Year = {2001},
}

@article{Freund01,
Author = {Freund, RW},
Title = {Computation of matrix-valued formally orthogonal polynomials and
   applications},
Journal = {J. Comput. Appl. Math.},
Year = {2001},
Volume = {127},
Number = {1-2, SI},
Pages = {173-199},
Month = {JAN 15},
DOI = {10.1016/S0377-0427(00)00505-7},
ISSN = {0377-0427},
EISSN = {1879-1778},
Unique-ID = {WOS:000166675200008},
}

@article{GoGue21,
Author = {Gosea, Ion Victor and Guttel, Stefan},
Title = {Algorithms for the rational approximation of matrix-valued
functions},
Journal = {SIAM J. Sci. Comput.},
Year = {2021},
Volume = {43},
Number = {5},
Pages = {A3033-A3054},
DOI = {10.1137/20M1324727},
ISSN = {1064-8275},
EISSN = {1095-7197},
ResearcherID-Numbers = {Güttel, Stefan/KBA-3474-2024},
Unique-ID = {WOS:000724956000001},
}

@article{Goychuk12,
   title={Viscoelastic subdiffusion: {G}eneralized {L}angevin equation approach},
   author={Goychuk, Igor},
   journal={Advances in Chemical Physics},
   volume={150},
   pages={187--253},
   year={2012},
   publisher={Wiley Online Library},
   DOI = {10.1002/9781118197714.ch5},
 }

@article{GKC09,
Author = {Gordon, Dan and Krishnamurthy, Vikram and Chung, Shin-Ho},
Title = {Generalized Langevin models of molecular dynamics simulations with
   applications to ion channels},
Journal = {J. Chem. Phys.},
Year = {2009},
Volume = {131},
Number = {13},
Month = {OCT 7},
DOI = {10.1063/1.3233945},
Article-Number = {134102},
Pages = {134102},
ISSN = {0021-9606},
EISSN = {1089-7690},
Unique-ID = {WOS:000270825500003},
}

@article{GS21,
Author = {Glatzel, Fabian and Schilling, Tanja},
Title = {The interplay between memory and potentials of mean force: A discussion
   on the structure of equations of motion for coarse-grained observables},
Journal = {EPL},
Year = {2021},
Volume = {136},
Number = {3},
Month = {NOV},
DOI = {10.1209/0295-5075/ac35ba},
Article-Number = {36001},
ISSN = {0295-5075},
EISSN = {1286-4854},
Unique-ID = {WOS:000761430900001},
}

@article{GLLB20,
Author = {Grogan, Francesca and Lei, Huan and Li, Xiantao and Baker, Nathan A.},
Title = {Data-driven molecular modeling with the generalized {L}angevin equation},
Journal = {J. Comp. Phys.},
Year = {2020},
Volume = {418},
Month = {OCT 1},
DOI = {10.1016/j.jcp.2020.109633},
Article-Number = {109633},
ISSN = {0021-9991},
EISSN = {1090-2716},
Unique-ID = {WOS:000561583600029},
}

@article{Hanke21,
Author = {Hanke, Martin},
Title = {Mathematical analysis of some iterative methods for the 
reconstruction of memory kernels},
Journal = {ETNA},
Year = {2021},
Volume = {54},
Pages = {483-498},
DOI = {10.1553/etna\_vol54s483},
ISSN = {1068-9613},
Unique-ID = {WOS:000715312600025},
}

@article{Hanke24,
Author = {Hanke, Martin},
Title = {Stochastic modeling of stationary scalar Gaussian processes in
   continuous time from autocorrelation data},
Journal = {Adv. Comp. Math.},
Year = {2024},
Volume = {50},
Number = {4},
Month = {AUG},
DOI = {10.1007/s10444-024-10150-7},
Article-Number = {60},
Pages = {60},
ISSN = {1019-7168},
EISSN = {1572-9044},
Unique-ID = {WOS:001253215700001},
}

@book{Higham08,
Author = {Higham, N. J.},
Title = {{Functions of Matrices: Theory and Computation, Algorithmic Differentiation}},
Publisher = {SIAM},
Year = {2008},
Address = {Philadelphia},
}

@article{JHS17,
 author = {Jung, Gerhard and Hanke, Martin and Schmid, Friederike},
 title = {Iterative Reconstruction of Memory Kernels},
 journal = {J. Chem. Theory Comp.},
 volume = {13},
 number = {6},
 pages = {2481-2488},
 year = {2017},
 doi = {10.1021/acs.jctc.7b00274},
 }

@article{JS21,
Author = {Jung, Gerhard and Schmid, Friederike},
Title = {Fluctuation-dissipation relations far from equilibrium: {A}  case study},
Journal = {Soft Matter},
Year = {2021},
Volume = {17},
Number = {26},
Pages = {6413-6425},
Month = {JUL 14},
DOI = {10.1039/d1sm00521a},
ISSN = {1744-683X},
EISSN = {1744-6848},
Unique-ID = {WOS:000661858900001},
}

@article{KTJSV21,
Author = {Klippenstein, Viktor and Tripathy, Madhusmita and Jung, Gerhard and
   Schmid, Friederike and van der Vegt, Nico F. A.},
Title = {Introducing Memory in Coarse-Grained Molecular Simulations},
Journal = {J. Phys. Chem. B},
Year = {2021},
Volume = {125},
Number = {19},
Pages = {4931-4954},
Month = {MAY 20},
DOI = {10.1021/acs.jpcb.1c01120},
ISSN = {1520-6106},
EISSN = {1520-5207},
Unique-ID = {WOS:000655615500003},
}

@article{KWV24,
Author = {Klippenstein, Viktor and Wolf, Niklas and van der Vegt, Nico F. A.},
Title = {A Gauss-Newton method for iterative optimization of memory kernels for
   generalized Langevin thermostats in coarse-grained molecular dynamics
   simulations},
Journal = {J. Chem. Phys.},
Year = {2024},
Volume = {160},
Number = {20},
Month = {MAY 28},
DOI = {10.1063/5.0203832},
Article-Number = {204115},
ISSN = {0021-9606},
EISSN = {1089-7690},
Unique-ID = {WOS:001235808600006},
}

@article{Kubo66,
     doi = {10.1088/0034-4885/29/1/306},
     year = 1966,
     month = {jan},
     publisher = {{IOP} Publishing},
     volume = {29},
     number = {1},
     pages = {255-284},
     author = {Kubo, R.},
     title = {The fluctuation-dissipation theorem},
     journal = {Rep. Progr. Phys.},
 }

@article{Laub,
     doi = {10.1088/0034-4885/29/1/306},
     year = {1979},
     volume = {24},
     pages = {913-921},
     author = {Laub, J.},
     title = {A {S}chur method for solving algebraic {R}iccati
     equations},
     doi = {10.1109/TAC.1979.1102178},
     journal = {IEEE Transactions on Automatic Control},
 }

@article{LAD19,
Author = {Lee, Hee Sun and Ahn, Suri-Hee and Darve, Eric F.},
Title = {The multi-dimensional generalized Langevin equation for conformational
   motion of proteins},
Journal = {J. Chem. Phys.},
Year = {2019},
Volume = {150},
Number = {17},
Month = {MAY 7},
DOI = {10.1063/1.5055573},
Article-Number = {174113},
Pages = {174113},
ISSN = {0021-9606},
EISSN = {1089-7690},
Unique-ID = {WOS:000467255500013},
}

@article {LBL16,
     author = {Lei, Huan and Baker, Nathan A. and Li, Xiantao},
     title = {Data-driven parameterization of the generalized Langevin equation},
     volume = {113},
     number = {50},
     pages = {14183-14188},
     year = {2016},
     doi = {10.1073/pnas.1609587113},
     publisher = {National Academy of Sciences},
     issn = {0027-8424},
     journal = {Proc. Natl. Acad. Sci. USA},
 }

@article{LLDK17,
 author = {Li, Zhen  and Lee, Hee Sun  and Darve, Eric  
    and Karniadakis,George Em },
 title = {Computing the non-Markovian coarse-grained interactions 
    derived from the {M}ori–{Z}wanzig formalism in molecular systems:
    {A}pplication to polymer melts},
 journal = {J. Chem. Phys.},
 volume = {146},
 number = {1},
 pages = {014104},
 year = {2017},
 doi = {10.1063/1.4973347},
 }

@article{LMPV22,
 author = {Lietaert, P. and Meerbergen, K. and P\'erez, J. and
 Vandereycken, B.},
 title = {Automatic rational approximation and linearization of
    nonlinear {E}igenvalue problems},
 journal = {IMA J. Numer. Anal.},
 volume = {42},
 pages = {1087-1115},
 year = {2022},
 doi = {10.1093/imanum/draa098},
 }

@article{MLL16a,
 author = {Ma, Lina  and Li, Xiantao  and Liu, Chun },
 title = {From generalized {L}angevin equations to {B}rownian dynamics 
   and embedded {B}rownian dynamics},
 journal = {J. Chem. Phys.},
 volume = {145},
 number = {},
 pages = {114102},
 year = {2016},
 doi = {10.1063/1.436761},
 }

@article{MLL16,
 author = {Ma, Lina  and Li, Xiantao  and Liu, Chun },
 title = {The derivation and approximation of coarse-grained dynamics
 from {L}angevin simulations},
 journal = {J. Chem. Phys.},
 volume = {145},
 number = {},
 pages = {204117},
 year = {2016},
 doi = {10.1063/1.4973347},
 }

@article{MG83,
 author = {Marchesoni,Fabio  and Grigolini,Paolo },
 title = {On the extension of the Kramers theory of chemical relaxation to the case of nonwhite noise},
 journal = {J. Chem. Phys.},
 volume = {78},
 number = {10},
 pages = {6287-6298},
 year = {1983},
 doi = {10.1063/1.444554},
 }

@article{MPS19,
 doi = {10.1209/0295-5075/128/40001},
 year = 2019,
 month = {jan},
 publisher = {{IOP} Publishing},
 volume = {128},
 number = {4},
 pages = {40001},
 author = {Meyer, Hugues and Pelagejcev, Philipp and Schilling, Tanja},
 title = {Non-Markovian out-of-equilibrium dynamics: 
     A general numerical procedure to construct time-dependent 
    memory kernels for coarse-grained observables},
 journal = {{EPL} (Europhysics Letters)},
 }

@article{Mori65,
     author = {Mori, Hazime},
     title = {{Transport, Collective Motion, and Brownian Motion}},
     journal = {Progr. Theor. Phys.},
     volume = {33},
     number = {3},
     pages = {423-455},
     year = {1965},
     month = {03},
     issn = {0033-068X},
     doi = {10.1143/PTP.33.423}
 }

@book{Pavliotis14,
Author = {Pavliotis, G. A.},
Title = {{Stochastic Processes and Applications: Diffusion Processes,
  the Fokker-Planck and Langevin Equations}},
Publisher = {Springer},
Year = {2014},
Address = {New York},
}

@article{HOOMD,
Author = {Anderson, Joshua A. and Glaser, Jens and Glotzer, Sharon C.},
Title = {HOOMD-blue: A Python package for high-performance molecular dynamics and
   hard particle Monte Carlo simulations},
Journal = {Computational Materials Science},
Year = {2020},
Volume = {173},
Month = {FEB 15},
DOI = {10.1016/j.commatsci.2019.109363},
Article-Number = {109363},
Pages = {109363},
ISSN = {0927-0256},
EISSN = {1879-0801},
Unique-ID = {WOS:000506172700065},
}

@article{NST18,
Author = {Nakatsukasa, Yuji and Sete, Olivier and Trefethen, Lloyd N.},
Title = {The {AAA} algorithm for rational approximation},
Journal = {SIAM J. Sci. Compt.},
Year = {2018},
Volume = {40},
Number = {3},
Pages = {A1494-A1522},
DOI = {10.1137/16M1106122},
ISSN = {1064-8275},
EISSN = {1095-7197},
Unique-ID = {WOS:000436986000033},
}

@article{Netz24,
Author = {Netz, Roland R.},
Title = {Derivation of the nonequilibrium generalized Langevin equation from a
   time-dependent many-body Hamiltonian},
Journal = {Phys. Rev. E},
Year = {2024},
Volume = {110},
Number = {1},
Month = {JUL 17},
DOI = {10.1103/PhysRevE.110.014123},
Article-Number = {014123},
Pages = {014123},
ISSN = {2470-0045},
EISSN = {2470-0053},
Unique-ID = {WOS:001279329700001},
}

@book{Risken_book,
 author = {Risken, H },
 title = {The Fokker-Planck equation},
 publisher = {Springer},
 address = {Berlin, Heidelberg, New York},
 year = {1996}
 }

@article{Schilling22,
Author = {Schilling, Tanja},
Title = {Coarse-grained modelling out of equilibrium},
Journal = {Phys. Rep.},
Year = {2022},
Volume = {972},
Pages = {1-45},
Month = {AUG 19},
DOI = {10.1016/j.physrep.2022.04.006},
ISSN = {0370-1573},
EISSN = {1873-6270},
Unique-ID = {WOS:000805191000001},
}

@article{SJS22,
Author = {Shea, Jeanine and Jung, Gerhard and Schmid, Friederike},
Title = {Passive probe particle in an active bath: can we tell it is out of
   equilibrium?},
Journal = {Soft Matter},
Year = {2022},
Volume = {18},
Number = {36},
Pages = {6965-6973},
Month = {SEP 21},
DOI = {10.1039/d2sm00905f},
ISSN = {1744-683X},
EISSN = {1744-6848},
Unique-ID = {WOS:000854083400001},
}

@article{SGTH10,
Author = {Siegle, Peter and Goychuk, Igor and Talkner, Peter and Haenggi, Peter},
Title = {Markovian embedding of non-Markovian superdiffusion},
Journal = {Phys. Rev. E},
Year = {2010},
Volume = {81},
Number = {1, 1},
Month = {JAN},
DOI = {10.1103/PhysRevE.81.011136},
Article-Number = {011136},
ISSN = {2470-0045},
EISSN = {2470-0053},
Unique-ID = {WOS:000274003300042},
}

@article{SKTL10,
 author = {Hyun Kyung Shin and Changho Kim and Peter Talkner and Eok Kyun Lee},
 title = {Brownian motion from molecular dynamics},
 journal = {Chemical Physics},
 volume = {375},
 number = {2},
 pages = {316 - 326},
 year = {2010},
 doi = {10.1016/j.chemphys.2010.05.019},
 }

@article{TDN24,
Author = {Tepper, Lucas and Dalton, Benjamin and Netz, Roland R.},
Title = {Accurate Memory Kernel Extraction from Discretized Time-Series Data},
Journal = {J. Chem. Theory Comp.},
Year = {2024},
Volume = {20},
Number = {8},
Pages = {3061-3068},
Month = {APR 11},
DOI = {10.1021/acs.jctc.3c01289},
ISSN = {1549-9618},
EISSN = {1549-9626},
Unique-ID = {WOS:001201286500001},
}

@article{VM22,
Author = {Vroylandt, Hadrien and Monmarche, Pierre},
Title = {Position-dependent memory kernel in generalized Langevin equations:
   Theory and numerical estimation},
Journal = {J. Chem. Phys.},
Year = {2022},
Volume = {156},
Number = {24},
Month = {JUN 28},
DOI = {10.1063/5.0094566},
Article-Number = {244105},
Pages={244105},
ISSN = {0021-9606},
EISSN = {1089-7690},
}

@article{WPDJ23,
Author = {Kerr Winter, Max and Pihlajamaa, Ilian and Debets, Vincent E. and
   Janssen, Liesbeth M. C.},
Title = {A deep learning approach to the measurement of 
   long-lived memory kernels from generalized {L}angevin dynamics},
Journal = {J. Chem. Phys.},
Year = {2023},
Volume = {158},
Number = {24},
Month = {JUN 28},
DOI = {10.1063/5.0149764},
Article-Number = {244115},
Pages = {244115},
ISSN = {0021-9606},
EISSN = {1089-7690},
Unique-ID = {WOS:001020762200002},
}

@article{WSW90,
Author = {Wang, Q. and Speyer, J. L. and Weiss, H.},
Title = {System characterization of positive real conditions},
Journal = {29th IEEE Conference on Decision and Control, Honolulu, USA},
Year = {1990},
Pages = {348-353},
}

@article{WMP20,
 author ={Wang, Shu and Ma, Zhan and Pan, Wenxiao},
 title  ={Data-driven coarse-grained modeling of polymers in solution 
    with structural and dynamic properties conserved},
 journal  ={Soft Matter},
 year  ={2020},
 volume  ={16},
 issue  ={36},
 pages  ={8330-8344},
 publisher  ={The Royal Society of Chemistry},
 doi  ={10.1039/D0SM01019G},
 }

@article{Zwanzig61,
 author = {Zwanzig, Robert },
 title = {Memory effects in irreversible thermodynamics},
 journal = {Physical Review},
 volume = {124},
 number = {4},
 pages = {983--992},
 year = {1961},
 doi = {10.1103/PhysRev.124.983}
 }

@book{Zwanzig_book,
 author = {Zwanzig, Robert },
 title = {Nonequilibrium statistical mechanics},
 publisher = {Oxford University Press},
 address = {New York},
 year = {2001}
 }

%%%%%%%%%%%%%%%%%%%%%%%%%%%%%%%%%%%%%%%%%%%%%%%%%%%%%%%%%%%

\end{document}